\documentstyle[12pt]{article}
\begin{document}
%
%


\textheight 9in
\topmargin -.25in
\oddsidemargin 0in
\evensidemargin 0in

 \jot=4mm
\textwidth=16truecm
\baselineskip=20pt    
\parindent=20pt
\footskip=48pt
\hoffset=-0.9cm  
\oddsidemargin=1cm 
\hsize=15.5truecm  
\setlength{\unitlength}{.1cm}
\pagenumbering{arabic}
\renewcommand{\theequation}{\thesection.\arabic{equation}}

\def\beqa{\begin{eqnarray}}
\def\eeqa{\end{eqnarray}}
\def\beq{\begin{equation}}
\def\eeq{\end{equation}}
\def\beqal{\begin{eqnarray}\label}
\def\beql{\begin{equation}\label}


\def\vol{\int d^4x\,\sqrt{-g}} 
\def\tvol{\int d^4x\,\sqrt{-\tilde{g}}} 
\def\grav{-\frac{1}{16 \pi }}
\def\half{\frac{1}{2}}
\def\gu{g^{\m\n}}
\def\gd{g_{\m\n}}
\def\tgd{\tilde{g}_{\m\n}}
\def\tgu{\tilde{g}^{\m\n}}
\def\lc{\raisebox{-.7ex}{$\stackrel{\textstyle <}{\sim}$}}
\def\gc{\raisebox{-.7ex}{$\stackrel{\textstyle >}{\sim}$}}
\def\R{\mbox{\rm I\kern-.18em R}}
\def\P{\mbox{\rm I\kern-.18em P}}
\def\Ds{\ {\big / \kern-.70em D}}
\def\uno{{1 \kern-.30em 1}}
\def\ds{\big / \kern-.90em {\ \p} }

\def\cDs{\ {\big / \kern-.70em {\cal D}}}
\def\Z{{Z \kern-.45em Z}}
\def\Q{{\kern .1em {\raise .47ex \hbox{$\scriptscriptstyle |$}}
\kern -.35em {\rm Q}}}
\def\de{\mbox{\it d}}
\def\Tr{\mbox{\rm Tr}}
\def\Im{\mbox{\rm Im}}
\def\cl{\mbox{\scriptstyle cl}}
\def\tr{\mbox{\rm tr}}
\def\ap{\alpha^{\prime}}
\def\psb{\bar{\psi}}
\def\chb{\bar{\chi}}
\def\lb{\bar{\lambda}}
\def\epsb{\bar{\varepsilon}}
\def\sb{\bar{\sigma}}
\def\ad{\dot{\alpha}}
\def\bd{\dot{\beta}}
\def\wb{\bar{w}}
\def\vb{\bar{v}}
\def\ib{\bar\imath}
\def\jb{\bar\jmath}
\def\um{^{\m}}
\def\un{^{\n}}
\def\dm{_{\m}}  
\def\dn{_{\n}}
\def\umn{^{\m\n}}
\def\dmn{_{\m\n}}
\def\umnrs{^{\m\n\r\s}}
\def\dmnrs{_{\m\n\r\s}}
\def\ua{^{\al}}  
\def\ub{^{\b}}
\def\da{_{\al}}
\def\db{_{\b}}
\def\ug{^{\g}}
\def\dg{_{\g}}
\def\uam{^{\al\m}}
\def\uan{^{\al\n}}
\def\uab{^{\al\b}}
\def\dab{_{\al\b}}
\def\dabgd{_{\al\b\g\d}}
\def\uabgd{^{\al\b\g\d}}
\def\udeab{^{;\al\b}}
\def\ddeab{_{;\al\b}}
\def\ddemunu{_{;\m\n}}
\def\udemunu{^{;\m\n}}
\def\ddemu{_{;\m}}  \def\udemu{^{;\m}}
\def\ddenu{_{;\n}}  \def\udenu{^{;\n}}
\def\ddea{_{;\al}}  \def\udea{^{;\al}}
\def\ddeb{_{;\b}}  \def\udeb{^{;\b}}

\def\naba{\nabla_{\a}}
\def\nabb{\nabla_{\b}}
\def\nabm{\nabla_{\m}}
\def\nabn{\nabla_{\n}}
\def\pmu{\partial_{\m}}
\def\pn{\partial_{\n}}
\def\p{\partial}

\def\bib#1{$^{\ref{#1}}$}

\def\al{\alpha} 
\def\b{\beta}
\def\g{\gamma}
\def\d{\delta}
\def\eps{\varepsilon}
\def\z{\zeta}
\def\h{\eta}
\def\th{\theta}
\def\k{\kappa}
\def\l{\lambda}
\def\m{\mu} 
\def\n{\nu} 
\def\x{\xi}
\def\r{\rho}
\def\s{\sigma}
\def\t{\tau}
\def\ph{\phi}
\def\ch{\chi}
\def\ps{\psi}
\def\om{\omega}
\def\G{\Gamma}
\def\D{\Delta}
\def\Th{\Theta}
\def\L{\Lambda}
\def\S{\Sigma}
\def\Ph{\Phi}
\def\Ps{\Psi}
\def\O{\Omega}

\def\ie{{\it i.e.}}
\def\cf{{\cal F}}
\def\ca{{\cal A}}
\def\cc{{\cal C}}
\def\cg{{\cal G}}
\def\cd{{\cal D}}
\def\cv{{\cal V}}
\def\cm{{\cal M}}
\def\co{{\cal O}}
\def\da{\dot{A}}
\def\db{\dot{B}}
\def\1{\dot{1}}
\def\2{\dot{2}}
\def\dd{\raisebox{+.3ex}{$\stackrel{\scriptstyle \leftrightarrow}{\partial}$}}
\def\etal{{\it et al.}\ }
\def\ie{{\it i.e. }}
\def\eg{{\it e.g. }}

\def\a{\`a }\def\o{\`o }\def\ii{\`\i{} }
\def\u{\`u  }\def\e{\`e }\def\ke{ch\'e }

\font\mybb=msbm10 at 12pt
\def\bb#1{\hbox{\mybb#1}}
\def\Z {\bb{Z}}
\def\R {\bb{R}}
\def\C {\bb{C}}
\def\H {\bb{H}}

\def\real{{\bb{R}}}
\def\rreal{{\bbb{R}}}
\def\rational{\bb{Q}}
\def\R4{\real^4}
\def\Ker{\mbox{Ker}}


\thispagestyle{empty}
\vskip 0.5cm
\begin{flushright}
{\tt hep-th/9706099}, 
ROM2F--97--26, DFPD 97/TH/26 \\
\end{flushright}
\centerline{\large \bf  NON--HOLOMORPHIC TERMS IN N=2 SUSY}

\vskip 0.2cm

\centerline{\large \bf  WILSONIAN ACTIONS AND RG EQUATION} 
\vspace{0.6cm}

\centerline{\sc 
D. BELLISAI$^1$, F. FUCITO$^{2}$, M. 
MATONE$^3$, 
G. TRAVAGLINI$^{2,4}$}
\vskip 0.2cm

\centerline{${}^1$ {\sl Dipartimento di Fisica, Universit\`{a} di Roma  ``Tor 
Vergata"}}
\centerline{{\sl Via della Ricerca Scientifica, 00133 Roma, ITALY}}
\vskip .2cm
\centerline{${}^2$ {\sl I.N.F.N. \ -- \ Sezione di Roma II ``Tor Vergata",
00133 Roma, ITALY}}
\vskip .2cm
\centerline{${}^3$ {\sl Dipartimento di Fisica ``G. Galilei" \ -- \  
I.N.F.N. \ -- \ 
Universit\`{a} di Padova}}
\centerline{{\sl Via Marzolo 8, 35131 Padova, ITALY}}
\vskip .2cm
\centerline{${}^4$ {\sl Dipartimento di Fisica, 
Universit\`{a} di Roma ``La Sapienza"}}
\centerline{{\sl P.le Aldo Moro 5, 00185 Roma, ITALY}}
\vskip 0.4cm

\centerline{\bf ABSTRACT}
{\noindent
In this paper we first investigate the Ansatz of one of the present authors for
$K(\Psi,\bar\Psi)$, the adimensional modular invariant non--holomorphic
correction to the Wilsonian effective Lagrangian of an $N=2$ globally 
supersymmetric gauge theory.
The  renormalisation group $\beta$--function of 
the theory crucially allows us to express 
$K(\Psi,\bar\Psi)$ in a form 
that easily generalises to the case in which the theory is 
coupled to $N_F$  hypermultiplets in the fundamental representation of 
the gauge group. 
This function satisfies an equation which should be viewed as a 
fully non--perturbative   
``non--chiral superconformal Ward identity". We also determine its 
renormalisation group equation. 
Furthermore, as a first step towards
checking the validity of this Ansatz, we compute the contribution to 
$K(\Psi,\bar\Psi)$ from  instantons 
of winding number $k=1$ and $k=2$.
As a by--product of our analysis we check a non--renormalisation theorem 
for $N_F=4$.}
\vskip .1in

\newpage
\textheight 8.5in
\baselineskip=24pt    
\setcounter{page}{1}
\setcounter{equation}{0}
\setcounter{section}{0}
\section{Introduction}
\indent In a celebrated paper,  Seiberg and 
Witten studied a globally $N=2$ supersymmetric 
Yang--Mills theory (SYM) with $SU(2)$ gauge group \cite{sw}. 
Subsequently they extended their analysis to theories with additional 
hypermultiplets (SQCD)\cite{sw2}.
They were able to exactly determine the
Wilsonian effective action up to 
two derivatives and four fermions 
In terms of an $N=2$ chiral superfield  $\Psi$, it 
is proportional to a holomorphic function ${\cal F}(\Psi)$
called the prepotential. From 
a physical point of view, the Wilsonian effective action 
describes the low--energy degrees of freedom 
of the $N=2$ microscopic supersymmetric theory. 
This achievement was possible thanks to a certain number of conjectures 
which
were suggested by the physics of the problem. It was later shown 
in \cite{mat2} that, in the case of $N=2$ SYM,  
these assumptions follow from the symmetries of the 
theory and from the inversion formula first derived in \cite{mat}
(subsequently generalised to SQCD in \cite{sonne}), 
and are consistent with microscopic
instanton computations in the cases of SYM
and SQCD \cite{fp,is,fp2,dkm,dkm2,DKMMATT,GAB}.

Since the moduli space of vacua of the theory is a 
thrice--punctured Riemann sphere, 
one can study the transformation properties of 
${\cal F}(\Psi)$ under
the modular group $\Gamma(2)$. The result of such an exercise is
the inversion formula in \cite{mat}, which   relates 
${\cal F}(\Psi)$ and its first derivative to a modular 
invariant function.  
The entire physical content of the theory can now be extracted
{}from this differential equation \cite{mat2,mat,dkm2,GAB}, 
which was also derived as an anomalous superconformal
Ward identity in \cite{HW}.

As it is well--known, a Wilsonian effective Lagrangian can be 
expanded in powers of the external momentum over some subtraction scale.
Much in the same vein of the previous analysis, 
the investigation of the modular
properties of the complete Wilsonian action leads to the conclusion
that the term with four--derivatives/8--fermions, which we 
will denote by $K(\Psi,\bar\Psi)$, is a modular 
invariant \cite{HENN}. 
However, it seems that also the higher--order terms are modular 
invariant. Let us denote the 
non--holomorphic part of the Wilsonian effective action by 
$\hat S[\Psi,\bar\Psi]$; furthermore, let $S, T$ be the $SL(2,\Z)$
generators with $S^2=1$ and $(ST)^3=1$.
In \cite{HENN}, it was shown that $\hat S[\Psi,\bar\Psi]$ does not
transform under the action of $T$ while, under duality,
$\cf(\Psi)\to{\cal F}_D(\Psi_D)={\cal F}(\Psi)+\Psi_D\Psi$. Now if
the action of $T$ on $\hat S[\Psi,\bar\Psi]$ is trivial and the 
group has only two generators, the action of $S$ must be trivial
too, since 
\beq
\hat S[\Psi,\bar\Psi]=
(ST)^3\circ\hat S[\Psi,\bar\Psi]=S^3\circ \hat S[\Psi,\bar\Psi]=
S\circ \hat S[\Psi,\bar\Psi]
\ \ .
\eeq
 However, we observe that the above modular invariance 
is considered with respect to the $S$ and $T$ action defined in
\cite{HENN} whereas, strictly speaking, a function $G(\Psi,\bar\Psi)$ is said 
to be modular invariant if 
$G(\gamma(\Psi),\gamma (\bar\Psi))=G(\Psi,\bar\Psi)$,
$\gamma\in SL(2,\Z)$.

Let us now leave this argument on the side and let us remark that
the perturbative 1--loop term and the contribution of instantons
of winding number $k=1$ to  $K(\Psi,\bar\Psi)$ were computed 
in \cite{DEWIT,YUNG}. On the basis of these results, and
by using uniformisation theory, one 
of the present authors was able to write a modular 
invariant function which satisfies the constraints imposed by 
perturbative and instanton calculations and which has no other 
singularities but the one at weak coupling \cite{MATK}.
This function satisfies the physical requirements of the theory,
for example it vanishes at those points of the moduli space where
monopoles or dyons become massless: we consider it to 
be a candidate for the expression of $K(\Psi,\bar\Psi)$. 
Its actual form will be reviewed in section 3 of this work,
where we also write it  
in terms of the $\beta$--function  of the theory, and
find the renormalisation group equation satisfied by $K$.
This function also an equation which should be viewed as a 
fully non--perturbative   
``non--chiral superconformal Ward identity". 
In that same section we also extend the Ansatz to the case of SQCD
with $N_F$ hypermultiplets. Furthermore, 
we  study the higher--derivative  corrections to the 
SYM and SQCD effective Lagrangians,  and in particular the 
contributions of instantons of winding number $k=1,2$ 
to the real adimensional function $K(\Psi , \bar{\Psi} )$. This is
a first step in the direction of checking the proposal
in \cite{MATK} and that of section 3.
As we will discuss in section 4,  
the situation is more involved than 
in the case of the holomorphic part of the effective Lagrangian, 
and we cannot provide here a check for the expression of 
$K(\Psi , \bar{\Psi} )$. 
We plan to come back on this point  in a future publication. 

The plan of the paper is the following:
in section 2 we briefly review the solution of \cite{sw} to
fix the notations and compute the relationship between the 
Pauli--Villars renormalisation group invariant scale and 
that appearing in \cite{sw}. We do this in great
detail because we will need it in the following and because 
the literature is plagued with inconsistent notations.
The content of section 3 has been discussed above. We start section 4 
by computing the $k=1$ contribution to  
$K(\Psi , \bar{\Psi} )$.
It turns out to be in agreement with the result of \cite{YUNG},
which was derived by different methods. In the second part of the
same section we compute the $k=2$ contribution, for $N=2$ SYM and 
SQCD. Furthermore,  we check a recent result 
concerning a non--renormalisation theorem in the case of four 
flavours \cite{SD}.
While we were writing this paper a work by Dorey {\it et al.}
\cite{DKMNEW} has appeared in which 
computations partly similar to ours, in the case of winding number $k=1$,
are carried out and the non--renormalisation theorem for 
$N_{F}=4$ is checked by using scaling arguments. 
Our results agree with theirs.

\section{A Review of the Seiberg--Witten Model}
\setcounter{equation}{0}
\indent
The Lagrangian density for the microscopic $N=2$ SYM theory, 
in the $N=2$ supersymmetric formalism is given by
\beq
\label{lagn=2n=2}
L= {1\over 16 \pi} \Im  \int \!  d^2 \theta d^{2} \tilde{\theta} \ 
\cf (\Psi ) \ \ .
\eeq
$\Psi$ transforms in the adjoint representation 
of the gauge group $G$ (which will be  $SU(2)$ from now on). 
Re--expressing the Lagrangian density in the $N=1$ formalism, 
we have 
\beq
\label{lagn=2}
L = {1\over 16 \pi}
\Im  \left[ \int \! d^2 \theta d^{2} \bar{\theta} 
K (\Phi , \bar\Phi , V ) + 
\int\! d^2 \theta f_{ab} (\Phi ) W^{a} W^{b} 
\right] \ \ ,
\eeq
where $a$, $b$ are indices of the adjoint representation of $G$.
The K\"{a}hler potential $K (\Phi , \bar\Phi , V ) $
and the holomorphic 
function  $f_{ab}(\Phi)$ 
are given, in terms of  $\cf$, by  
\beq
 K (\Phi , \bar\Phi , V )  =   
(\bar\Phi e^{-2V})^{a} {\p\cf \over \p \Phi^a}
\ \ ,
\eeq
\beq
 f_{ab} (\Phi) =  {\p^2 \cf \over \p \Phi^a \p\Phi^b}  
\ \ .
\eeq
The classical action for the $N=2$ SYM  theory   
is obtained by choosing for $\cf$ 
the functional form 
\beq
\label{preclass1}
\cf_{\rm cl} ( \Psi ) = { \tau_{\rm cl} \over 2} (\Psi^a \Psi^a  )\ \ ,
\eeq
where we conventionally define $\tau_{\rm cl}$  as
\beq
\label{preclass2}
\tau_{\rm cl}= {\theta \over 2\pi} + {4\pi i \over g^2} \ \ .
\eeq
Our normalisations are the same as in \cite{sw}.
After eliminating the auxiliary fields, 
the classical action of the theory is given by
\beq
S =S_{\rm G} +  S_{\rm H} +  S_{\rm F} + S_{\rm Y}+ S_{\rm pot}
\ \ .
\label{azione}
\eeq 
$S_{\rm G}$ is the usual gauge field action, 
the kinetic terms for the Fermi and Bose fields  
minimally coupled to the gauge field $A_{\mu}$ are 
\beq
S_{\rm F}[\lambda,\bar\lambda,A]= \int d^{4}x\ 
\bar{\lambda}^{\dot{A} a } \Bigl[ \ 
\Ds(A)\lambda_{\dot{A}} \Bigr]^{a}
\ \ ,
\eeq
where $\l_{\dot{A}}$ are the two gauginos, 
$\dot{A}=1,2$, and   
\beq
S_{\rm H}[\phi,\phi^\dagger,A]=\int d^{4}x\ (D \phi)^{\dagger a}
(D \phi)^{a} 
\ \ .
\eeq
The Yukawa interactions are given by
\beq
S_{\rm Y} [ \phi , \phi^{\dagger} , \lambda , \bar{\lambda}] = 
\sqrt{2} g  \epsilon^{abc} 
\int d^4x \ \phi^{a \dagger} (\lambda_{\dot{1}}^{b} 
\lambda_{\dot{2}}^{c}) \ + \ \mbox {\rm h.}\mbox{\rm c.}
\eeq
and finally
$S_{pot} =  \int d^{4} x\  V(\phi,\phi^{\dagger})$
comes from the potential term  
\beq
\label{anmm.1}
V (\phi , \phi^{\dagger}) = 
\Tr  [ \phi , \phi^\dagger ]^2 
\ \ ,
\eeq
for the complex scalar field. 
As required by supersymmetry, one has
$V(\phi , \phi^{\dagger})\geq 0$.
The condition $V(\ph , \ph^{\dagger}) = 0$ implies that 
$[\ph , \ph^{\dagger} ]=0$: 
$\ph$ is then a normal operator, and can be diagonalised 
by a unitary matrix: that is, a colour rotation.
The most general (supersymmetric) classical vacuum configuration 
is then   
\beq
\label{vaspf}
\ph_{0} = a \left( \Omega {\sigma_3 \over 2}\Omega^{\dagger} 
\right) \ \ , 
\ a \in \C \ \ , \ \ \Omega \in SU(2) \ \ .
\eeq
When $a \neq 0$ the $SU(2)$ gauge symmetry is spontaneously broken 
to $U(1)$.
The classical vacuum ``degeneracy''  for the  $N=2$ SYM 
theory is lifted neither by perturbative nor by non--perturbative
quantum corrections \cite{Seibergnonren,gomez}. 
In fact any non--zero superpotential    would explicitly break 
the extended supersymmetry of the model;
however the Witten index of the theory is non--zero \cite{Witten1},
so supersymmetry stays unbroken. We then have a 
full quantum moduli space,   $\cm_{SU(2)}$,  for the low--energy 
theory. 
The effective Lagrangian for the massless $U(1)$ fields  
 $\Phi = \left\{ \phi^{3}, \l_{\alpha \1}^{3}, F^3 \right\}$
will again be of the form
\beq
\label{lagmassless}
L_{eff} = 
{1\over 16 \pi}
\Im  \left[ \int\! d^2 \theta \cf^{\prime \prime} (\Phi ) W  W + 
\int \! d^2 \theta d^{2} \bar{\theta} 
\bar\Phi \cf^{\prime} (\Phi )   
\right] \ \ ,
\eeq
where 
$W= \left\{ A_{\mu}^{3} ,\l_{\alpha \2}^{3}, D^3
\right\}$.

\noindent
The low energy dynamics is then governed by 
a unique  function 
$\cf (\Phi )$, the effective prepotential, 
whose functional form is not restricted by supersymmetry.
The crucial property of $\cf (\Phi )$, first proved in 
\cite{Gates}, is holomorphicity.
In analogy with (\ref{preclass2})  we can also define an effective coupling 
constant as
\beq
\label{ol.4}
\tau (a ) = \cf^{\prime \prime} (a ) \ \ .
\eeq

\noindent
It is a simple exercise to rewrite 
(\ref{lagmassless}) in the component field formalism.
\footnote{Throughout 
the article we will use the conventions of Wess and Bagger 
\cite{WB} for the product of Weyl spinors and integration on superspace.
We also define the Euclidean $\s_{\m}$,  $\sb_{\m}$ matrices as 
$\s_{\m} = (\uno , i \s^{a} )$,
$
\sb_{\m} = (\uno , -  i \s^{a} ) 
$,
$\s^{a}$, $a=1,2,3$ being the usual Pauli matrices, and the (anti)self-dual 
matrices $(\sb_{\m\n})\, \s_{\m\n}$ are 
$
\s_{\m \n}={1\over
2}(\s_{\m}\sb_{\n}-\s_{\n}\sb_{\m})=i\eta^{a}_{\m \n} \s^{a}$,
$
\sb_{\m\n}={1\over 2}(\sb_{\m}\s_{\n}-\sb_{\n}\s_{\m})=
i\bar{\eta}^{a}_{\m \n} \s^{a}
$, 
where $\eta^{a}_{\m\n},\bar{\eta}^{a}_{\m\n}$ are the 't Hooft symbols 
defined in \cite{th}.}
This way we obtain
\beqa
\label{ol.3}
L_{eff}  &= &
{1\over 4 \pi}
\Im  \biggl[ 
- \cf^{''} (\ph) \left( |\p_\mu \ph |^2 + i \lb_{\dot{A}}\ \ds  
\l^{\dot{A}}
 + {1\over 4} F_{\mu \nu} F_{\mu \nu}
\right) +  
\nonumber \\
&&
{1\over\sqrt{2}} \cf^{'''} (\ph) \l_{\1} \sigma^{\mu \nu}\l_{\2} 
F_{\mu \nu} + {1\over 4} \cf^{IV} (\ph )  \l_{\1}^2\l_{\2}^2 \biggr] 
+  \cdots
\ \ ,
\eeqa
where the dots stand for 
terms of higher order in the coupling constant.
The effective description of the low--energy dynamics in terms of the 
$U(1)$ superfields $\Phi$, $W$ is not appropriate for all vacuum configurations.
In particular, the quantum moduli space  $\cm_{SU(2)}$
is better described in terms of the variable $a$ and its 
dual $a_{D}=\p_{a} {\cal F}$. When the gauge group is 
$SU(2)$, we can describe $\cm_{SU(2)}$
in terms of the 
gauge--invariant coordinate $u = <\Tr \phi^2 >$.   Then $\cm_{SU(2)}$
is the Riemann sphere with punctures at $u=\infty$
and, in the normalisation of \cite{sw}, at $u=\pm \Lambda^2$.\\ 
At the classical level
\beq
\cf_{\rm cl} ( a )  = { \tau_{\rm cl} \over 2} a^2 \ \ ,
\eeq
however 
perturbative as well as non--perturbative effects modify the 
expression of the prepotential.
We shall then write
\beq
\cf (a) = \cf_{pert} (a)  +\cf_{np} (a) 
\ \ ,
\eeq
including the classical contribution in the first term.
The perturbative contribution has been calculated by Seiberg  
\cite{sei}
and is 
 exactly determined thanks to the 
holomorphicity requirements on $\cf (a)$ 
and to the $U(1)_R$
symmetry 
\beq
\label{u1r}
U(1)_R: \ \ \l_{\da}\longrightarrow e^{i\alpha} \l_{\da}\ , \ \ 
\ph \longrightarrow e^{2i\alpha} \ph \ .
\eeq
The associated current $J^{\mu}_{R}$  is anomalous
\beq
\label{Mozart}
J_{\mu R} = 
\lb_{\1 }\sb_{\mu}\l_{\1} + \lb_{\2} \sb_{\mu}\l_{\2}  + 
2 i \ph^{\dagger}\dd_{\!\mu}\ph
\ \ , \ 
\p_{\mu} J^{\mu}_{R} =  - \frac{i}{32 \pi^2}
(F_{\mu \nu}^{a} \tilde{F}_{\mu \nu}^{a} )  (4 N_c )
\ \ ,
\eeq
(in our case the number of colours is taken to be $N_c = 2$).
The discrete subgroup $\Z_{8}\subset U(1)_R$,
generated by the transformations 
(\ref{u1r}) with $\alpha_{m}=(2\pi / 8 ) m$, $m\in \Z $
is a symmetry of the full quantum theory, since in this case 
the action functional $S$ transforms as 
\beq\label{ciccillo}
S\longrightarrow S + i8  k  \alpha_m= S + 2\pi i m \ \ .
\eeq
At a given point in the $u$--moduli space the $\Z_{8}$ symmetry 
spontaneously breaks down to $\Z_{4}$, since the $U(1)_{R}$ charge of $u$ 
is +4. However, (\ref{ciccillo}) tells us that the points $u$ and $-u$ 
correspond to physically equivalent theories. 
We now immediately rewrite  (\ref{u1r}) 
in terms of the $U(1)$ superfield $\Psi$ of the $N=2$ supersymmetry
as
\beq
\label{u1rn=2}
U(1)_R: \ \ \Psi (x, \theta) \longrightarrow 
\Psi^{\prime} (x, \theta^{\prime})=
e^{2i\alpha} \Psi (x, \theta  e^{-i\alpha}) \ \  :
\eeq
if we now assign a charge of $+1$ to  $\theta$, 
the charge of $\Psi$ will be $+2$ 
in such a way that the classical prepotential (\ref{preclass1})   
is invariant. 
Then the perturbative effective Lagrangian
\beq
\label{lagn=2n=2pert}
L_{pert} [ \Psi ]
= {1\over 16 \pi} {\Im}  \int \!  d^2 \theta d^{2} \tilde{\theta} \ 
\cf_{pert} [\Psi (x, \theta )] \ \ ,
\eeq
transforms in 
\beqa
L_{pert}^{(\alpha )} [ \Psi^{\prime} ]
&=& {1\over 16 \pi} \Im  \int \!  d^2 \theta d^{2} \tilde{\theta} \ 
\cf_{pert} [e^{2i\alpha} \Psi (x, \theta e^{-i\alpha} ) ] = 
\nonumber \\
&& 
{1\over 16 \pi} \Im  \int \!  d^4 \theta   \ 
e^{-4i\alpha}\cf_{pert} [e^{2i\alpha} \Psi (x, \theta   ) ]
\ \ ,
\eeqa
where $d^4 \theta = d^2 \theta d^{2} \tilde{\theta}$.
After a little algebra we get
\beq
\label{488}
L_{pert}+ \delta_{\alpha}L_{pert}= 
{1\over 16 \pi} \Im  \int \!  d^4 \theta   \left[
 1 + 4i\alpha \left( -1  +
 \Psi^2 {\p \over
\p \Psi^2 } \right)\right]\cf_{pert} (\Psi )   
\ .
\eeq
Furthermore, we know that under a 
$U(1)_R$ transformation
\beq
\delta_{\alpha}L_{pert}= -(4 N_c \alpha) 
\left( \frac{1}{32 \pi^2}
F_{\mu \nu}^{a} \tilde{F}_{\mu \nu}^{a} \right) \ \ ,
\eeq
(with  $N_c=2$), so that 
\beq
\label{suona}
L_{pert}+ \delta_{\alpha}L_{pert}= 
{1\over 16 \pi} \Im  \int \!  d^4 \theta \ 
\left[ \cf (\Psi ) - {2\alpha \over \pi }\Psi^2
\right] \ \ ,
\eeq
{}from which it immediately follows that
\footnote{Disregarding terms which vanish when integrated 
in $d^4 \theta$.}
\beq
\label{ac.1}
 (\Psi^2 {\p  \over\p \Psi^2 }-1)\cf_{pert} (\Psi ) 
 = {i\over 2\pi}\Psi^2 
\ \ .
\eeq
This is the semiclassical version
\cite{HW,gt,algom} of the non--perturbative relation 
\beq
\label{f.2}
i\pi\left(\cf-{a\over 2}
{\partial\cf\over\partial a}\right)= <\Tr\ \phi^2>
\ \ ,
\eeq
obtained in \cite{mat} and subsequently re--derived in \cite{HW}.
The solution of (\ref{ac.1}) is
\beq
\cf_{pert}  (\Psi ) =  {i\over 2\pi}\Psi^2  \ln 
{\Psi^2 \over \mu^2} \ \ ,
\eeq
where $\mu$ can be fixed   
by the value of  the coupling constant at some subtraction 
point. 
The normalisation of the (one--loop)  perturbative contribution 
must be fixed together with the non--perturbative contributions 
and the definition of the renormalisation group invariant (RGI) scale 
$\Lambda$.
To this end  we first write the non--perturbative prepotential as
\beq
\label{nat.01}
\cf_{np} (a) = 
\sum_{k=1}^{\infty} \cf_k \left( {\Lambda \over a }\right)^{4 k}  a^2 
\ \ ,
\eeq
and similarly
\beq
\label{espditrfiquad}
u(a) ={1\over 2} a^2 +  \sum_{k=1}^{\infty} \cg_k 
\left({\Lambda\over a}\right)^{4k } a^2 
\ \ . 
\eeq
It easy to check that the expressions (\ref{nat.01}), (\ref{espditrfiquad})
possess the correct invariance  properties under the ${\Z}_{8}$ symmetry.
The values of the $\cf_k$'s and the  $\cg_k$'s
are meaningful only if 
one specifies the choice of the RGI scale $\Lambda$, 
and  can be obtained 
via a $k$--instanton calculation
\cite{fp,dkm,GAB}. 
In the following we shall need the expressions for the 
1--instanton contributions to $u(a)$, which has been found to be 
\cite{fp,GAB}
\beq
\label{cg1pv}
<\Tr \phi^2 >_{k=1} = {\Lambda_{PV}^4 \over a^2}
\ \ .
\eeq
Here 
$\Lambda_{PV}$ is the Pauli--Villars RGI invariant scale, 
which naturally arises when performing instanton calculations
after the cancellation of the determinants of the kinetic operators 
of the various fields \cite{divecchia}. 
We shall fix
in a moment its relationship with 
the scale employed in \cite{sw}.
Note that the relation (\ref{f.2}) gives 
the $\cf_k$'s as a function of the $\cg_k$'s, 
\beq 
\label{ca.13}
2 i \pi k \cf_k = \cg_k 
\ \ .
\eeq
By making some hypotheses on the structure of
the moduli space and on the monodromies of $\tau$ around its  singularities, 
Seiberg and Witten were able to obtain the expressions of $a(u)$ and 
$a_{D}(u)$, which are given by 
\beqa
\label{adiu}
a(u)&=&\frac{\sqrt{2}}{\pi}\int_{-\L^{2}}^{\L^{2}} d x\ \frac{\sqrt{x-u}}
{\sqrt{x^2-\L^{4}}}
\ \ ,
\\
\label{addiu}
a_{D}(u)&=& \frac{\sqrt{2}}{\pi}
\int_{\L^{2}}^{u} dx \ \frac{\sqrt{x-u}}
{\sqrt{x^2-\L^{4}}}
\ \ ,
\eeqa
where $\Lambda$ is the Seiberg--Witten RGI  scale (to 
be matched against the  Pauli--Villars one).
We now put 
\beq
a_D (u) = {\sqrt{2u}\over \pi} g( 1 / u)
\ \ ,
\eeq
where 
\beqa
g(1 / u) &=&
\int_{\L^{2} / u}^{1} d z\ \frac{\sqrt{z-1}}
{\sqrt{z^2-\L^{4} / u^2}} = 
\noindent \\
& &\int_{\L^{2} / u}^{1} d z\  \biggl[
\frac{\sqrt{z-1}}{\sqrt{z^2-\L^{4} / u^2}}
- {i\over z}\biggr]
 + 
\int_{\L^{2} / u}^{1} d z\ 
{i\over z} 
\ \ .
\eeqa
The perturbative constant ($ u \gg \Lambda^2  $) contribution to 
$g(1 / u )$ is
\beq
\int_{0}^{1} d z\  \biggl[
\frac{\sqrt{z-1}}{z}
- {i\over z}\biggr] = 2 i \ln {2\over e}
\ \ ,
\eeq
so that 
\beq
a_D (u) \to 
i {\sqrt{2u}\over \pi} \ln {4 u\over (e \Lambda )^2} 
\ \ .
\eeq
Using the asymptotic expansion
(\ref{espditrfiquad}) we finally obtain an expression for $a_D$ as a function 
of $a$ in the perturbative regime, 
\beq
a_D (a) \to {i\over \pi}a \ln {2 a^2 \over (e\Lambda )^2}
\ \ ,
\eeq
that is, in the same limit, 
\beq
\label{ciccio}
\cf_{pert} (a) =
 {i\over 2\pi}a^2 \ln {2 a^2 \over e^3\Lambda^2}
\ \ .
\eeq
This sets the normalisation of the classical and perturbative contributions.
{}From (\ref{ciccio}) it follows that 
\beq
\tau_{pert} (a) = \cf_{pert}^{\prime \prime} (a ) = 
 {i\over \pi} \ln {2 a^2 \over \Lambda^2}
\ \ .
\eeq
We now examine the first instanton correction to $u(a)$; via the relation
(\ref{ca.13}) we will then fix the normalisation of the $\cf_{k}$'s.
Expanding the expression (\ref{adiu}) 
of the modulus $a$ as a function of $u$ for $ u \gg \Lambda^2  $
we get
\beqa
\label{car.77}
a(u) &= &
\frac{\sqrt{2}}{\pi}\biggl[
\int_{-\L^{2}}^{\L^{2}} d x\ \frac{1}
{\sqrt{\L^{4}-x^2}} - 
{1\over 8 u^2} \int_{-\L^{2}}^{\L^{2}} d x\ \frac{x^2}
{\sqrt{\L^{4}-x^2}} 
+ \mbox{{\rm  O}}(\L^8 / u^4 )
\biggr] = 
\nonumber \\
& &
\sqrt{2u} \Bigl[ 1 - {\L^4 \over 16 u^2} + 
\mbox{{\rm O}}(\L^8 / u^4 )
\Bigr]
\ \ . 
\eeqa
In the same approximation we also have that
\beq
u(a) = {a^2 \over 2} \left[ 1 + 2 \cg_1 \left( {\L \over a } \right)^4 + 
\mbox{{\rm  O}}\left( {\L \over a}\right)^8
\right]
\ \ ;
\eeq
substituting into (\ref{car.77}) we get
\beq
a=
a \left[ 1 +  \cg_1 \left( {\L \over a } \right)^4 + 
\mbox{{\rm  O}}\left( {\L \over a}\right)^8
\right]\cdot
\left\{ 1- {1 \over 4}\left( {\L \over a} \right)^4\left[
1 + \mbox{{\rm  O}}\left( {\L \over a}\right)^8
\right] \right\}
\ \ ,
\eeq
and,  for consistency, we must impose 
\beq
\label{cg1sw}
\cg_1 = {1\over 4}\ \ ,
\eeq
with respect to the RGI scale $\L$ in \cite{sw}.
Comparing
(\ref{cg1sw}) with (\ref{cg1pv})
we find
\beq
\L_{PV} = {\L \over \sqrt 2}
\ \ .
\eeq
The holomorphic prepotential ${\cal F}(a)$
is then given by
\beqa
{\cal F}(a)&=&\frac{i}{2\pi}a^{2}\ln\frac{2a^2}{e^3 \L^2}+
a^2 \sum_{k=1}^{\infty}{\cal F}_{k}\left(\frac{\L}{a}\right)^{4k}= 
\nonumber \\
& & 
\frac{i}{2\pi}a^{2}\ln\frac{a^2}{e^3 \L_{PV}^2}+
a^2 \sum_{k=1}^{\infty}{\cal F}_{k} \ 2^{2k} \left(\frac{\L_{PV}}{a}\right)^{4k}
\ \ ,
\eeqa
where ${\cal F}_{1}= \cg_1/   2\pi i  = 
1 /  8\pi i$.
Finally, when we add 
$N_{F}$ hypermultiplets, the holomorphic prepotential ${\cal F}^{(N_{F})}(a)$ 
becomes 
\beq
{\cal F}^{(N_{F})}(a)=
\frac{i}{8\pi}(4-N_{F})a^{2}\ln\frac{a^2}{e^3 (\L_{PV}^{(N_{F})})^2}+
a^2 \sum_{k=1}^{\infty}{\cal F}^{(N_{F})}_{k} \ 2^{2k} 
\left(\frac{\L_{PV}^{(N_{F})}}{a}\right)^{k(4-N_{F})}
\ \ ,
\eeq
where ${\cal F}^{(N_{F})}_{2k+1}$=0 in the presence of 
massless hypermultiplets. This 
is a consequence of the $Z_{4(4-N_{F})}$ chiral symmetry group 
of the full quantum theory \cite{sw2}. 
\section{Non--holomorphic corrections and the \boldmath${\b}$--function}
\setcounter{equation}{0}
Let us briefly describe the general form of 
the higher--derivative corrections to the 
Lagrangian (\ref{lagn=2}). 
Since an effective Lagrangian is written as 
an expansion in the space of momenta, the next--to--leading 
contributions will come out of the terms with four or more derivatives or 
eight or more fermions. In the case of $N=2$ SYM theory, they will be 
written as a finite expansion in spinor derivatives,  
\beq
\label{kappasnl}
S_{NL}(\Psi,\bar{\Psi})=\int d^4 x d^4 \th d^4 \bar{\th}
\left[ 
K(\Psi,\bar{\Psi})+
G(\Psi,\bar{\Psi})D\Psi D\Psi \bar{D}\bar{\Psi}\bar{D}\bar{\Psi}+\cdots+
{\mbox{\rm O}}(D^4,\bar{D}^4)\right]
\ \ .
\eeq
If we assign the scaling dimension 
$[dx]=1$, $[d\th]=-1/2$ 
and $[D]=1/2$, 
as a consequence of $N=2$ supersymmetry the expansion will contain only 
terms with  even dimension.
Furthermore,  the $U(1)_{R}$ anomaly and the non--perturbative 
corrections are completely 
encoded in the analytic prepotential $\cf$,
which is the only holomorphic term that can appear in the effective 
Lagrangian.
Therefore (\ref{kappasnl})
is integrated over the whole superspace. 
{}From now on we will restrict
our attention  to the first term
$K(\Psi,\bar{\Psi})$ in  (\ref{kappasnl}),  which is adimensional
and does not contain 
spinor derivatives of $\Psi$ and $\bar{\Psi}$.

We now consider the derivation of
$K$ proposed in \cite{MATK}.
Let $H=\{w|\Im\, 
w>0\}$ be the upper half plane endowed with the 
Poincar\'e metric $ds^2_P=(\Im\, w)^{-2}|dw|^2$. Since $\tau=\partial_a^2 
{\cal F}$ is the inverse of the map uniformising ${\cal M}_{SU(2)}$,
it follows that the positive--definite  metric
\begin{equation}
ds_P^2={\left|{\partial^3_a{\cal F}}\right|^2\over\left(\Im\, 
\tau\right)^2}|da|^2={\left|\partial_u \tau\right|^2\over\left(\Im\,
\tau\right)^2}|du|^2=e^{\varphi} |du|^2
\ \ ,
\label{16bb}
\end{equation}
is the Poincar\'e metric on ${\cal M}_{SU(2)}$. This implies that $\varphi$ 
satisfies the Liouville equation 
\begin{equation}
\partial_{\bar u}\partial_u\varphi={e^{\varphi}\over 2}
\ \ .
\label{opop1j}
\end{equation}
Observe that this equation is satisfied since,  
for any fundamental
domain $F$ in $H$,  $\tau(u)$ is a {\it 
univalent} ({\it i.e.} one--to--one) map between ${\cal M}_{SU(2)}$ and $F$.
 In this context we stress that $\tau(u)$ 
is not properly a function; rather it is a {\it polymorphic} function
({\it i.e.} it is M\"obius transformed after going around
non--trivial cycles).
Therefore classical theorems concerning standard 
meromorphic functions do not hold. In 
particular,  $\Im\, \tau(u)$ is a zero mode of the
Laplacian. Observe that on 
the moduli space $\tau(u)$ is holomorphic as zeroes and poles
are at the punctures (that is missing points).
Zeroes and poles are manifest on the compactified moduli space.
However, these critical points are absent in the case of higher genus 
Riemann surfaces without punctures. This follows from the fact that
punctures correspond to points $\tau\in \real=\partial H$. In 
particular, as the fundamental domains of negatively curved Riemann 
surfaces without punctures $\Sigma$ belong
to $H$, it follows that in these cases
$\tau$ is a holomorphic nowhere vanishing function
on $\Sigma$. In particular, $\Delta \ \Im\,\tau=0$.
In \cite{mat,m2,mat2} it was shown how the results of \cite{sw} 
are naturally described in the framework of uniformisation theory.
We now show how the function $K(a,\bar a)$ derived in
\cite{MATK} naturally arises in this 
context.

To see this let us first recall some asymptotics
for the Poincar\'e metric. 
Let us consider the Riemann sphere
with elliptic or parabolic  points (punctures)
at $u_1,...,u_{n-1}$, $u_n=\infty$.
Near an elliptic point the behaviour of the Poincar\'e metric is
\begin{equation}
e^{\varphi}\sim{4q^2_k
r_k^{2q_k-2}\over\left(1-r_k^{2q_k}\right)^2}\ \ ,
\label{bhv3}
\end{equation}
where $q_k^{-1}$ is the ramification index of $u_k$ and 
$r_k=|u-u_k|$, $k=1,\ldots, n-1$, $r_n=|u|$. Taking the $q_k\to 0$
limit we get the parabolic singularity (puncture)
\begin{equation}
e^{\varphi}\sim {1\over r_k^{2}\log^2r_k}\ \ .
\label{bhv2}
\end{equation}
It follows that
in ${\cal M}_{SU(2)}$ case the 
Poincar\'e metric 
$e^\varphi$ vanishes only at the puncture $u=\infty$, where 
$\varphi\sim -2\ln(|u|\ln|u|)$. 
Furthermore, $e^{\varphi}$ is divergent only 
at the punctures 
$u=\pm\Lambda^2$, where $\varphi\sim -2\ln(|u\mp\Lambda^2|\ln|u
\mp\Lambda^2|)$.

Let us now gather the known results on $K(\Psi, \bar{\Psi})$.
First observe that in \cite{DEWIT} it was proved that, 
to the one--loop order 
\begin{equation}
{K}(\Psi,\bar{\Psi})\sim c\ln {\Psi\over \Lambda}\ln {\bar{\Psi}
\over\Lambda}\ \ ,
\label{hasthestr}
\end{equation}
where $c$ is a constant 
which was recently calculated  \cite{ketov}  
in the formalism of harmonic superspace
for $0\leq N_F \leq 4$.  
The non--holomorphic terms in the effective 
Lagrangian are $U(1)_{R}$--invariant. 
If we follow the reasoning made for $\cf$ (which eventually led 
to (\ref{ac.1})) we get, in particular, 
\beq\label{muh}
\int d^4 \th d^4 \bar\th \left\{\Psi\frac{\p}{\p \Psi}-\bar\Psi
\frac{\p}{\p \bar\Psi}\right\}K(\Psi,\bar\Psi)=0\ \ ,
\eeq
which should be considered as a semiclassical Ward identity for $K$. 
The solution of this equation is simply given, modulo
K\"ahler transformations, by
\beq
\label{nostra}
K(y,\bar{y})=P(y+\bar{y})+y \bar{g}(\bar{y})+\bar{y} g(y)
\ \ ,
\eeq
where $y=\ln \Psi/\L$, 
$g$ is an arbitrary function and $P=\bar{P}$.
\footnote{It is a trivial exercise to show that (\ref{muh}) 
is completely equivalent to the superspace--integrated version of 
equation (3.7) of \cite{ketov}.}
In particular, the term found in 
\cite{DEWIT} is a solution to this equation,  but it seems that, 
in principle, no 
non--renormalisation theorem 
prevents us from considering solutions with higher order polynomials 
in $\ln(\Psi\bar{\Psi}/\L^2)$. These terms would 
represent higher--loop contributions to $K$.  
However, in the case of SQCD with $N_{F}=4$ massless 
hypermultiplets and gauge group $SU(2)$,  we know 
that the $\b$--function vanishes, so that no scale $\L$ 
arises in the theory. 
In this case the only possible function of $\Psi/\L$ which can appear 
in the solution (\ref{nostra}) is a term linear 
in the product $\ln(\Psi/\L) \ln(\bar\Psi/\L)$ (or, up to 
purely chiral or antichiral terms, quadratic in $\ln(\Psi\bar{\Psi}/\L^2)$) 
\cite{SD};
indeed, only in this case the scale $\L$ is a fake (it  does not multiply 
non--holomorphic terms in the Lagrangian),  as it 
should for a scale--invariant theory. 

Let us go back to the $N_F=0$ case. Besides (\ref{hasthestr})
we know that
that ${K}$ is a modular  invariant \cite{HENN}
and that the one--instanton  contribution is \cite{YUNG}
\footnote{From now on we will denote by $\Lambda$ the Pauli--Villars
RGI scale.}
\begin{equation}
{K}(\Psi,\bar \Psi)|_{k=1}= 
{1\over 8 \pi^2}\left({\Lambda\over\Psi}\right)^{4}
\ln {\Psi\over \Lambda} {\bar \Psi\over \Lambda}+
{\mbox {\rm h.}}{\mbox{\rm c.}}
\label{oneYung}
\end{equation}
Strictly speaking, a 
function $G(\Psi,\bar\Psi)$ is said to be modular invariant if
$G(\gamma(\Psi),\gamma (\bar \Psi))=G(\Psi,\bar \Psi)$, 
$\gamma\in SL(2,{\Z})$.
However, 
${K}(\Psi,\bar \Psi)$ has the invariance $T\circ {K}(\Psi,\bar \Psi)
={K}(\Psi,\bar \Psi)$ and $S\circ { K}(\Psi,\bar \Psi)= K(\Psi,\bar \Psi)$. 
While  in the former case there is no  change in the functional structure 
of ${K}$, in the latter, according to the 
$S$--dual formulation of the theory, where
${\cal F}(\Psi)$ is replaced by ${\cal F}_D(\Psi_D)$, 
the function  
$S\circ {K}(\Psi,\bar  \Psi)$ 
should be constructed  with the building block  ${\cal F}_D(\Psi_D)$
(which replaces ${\cal F}(\Psi)$
in the construction of ${K}(\Psi,\bar \Psi)$).

Let us discuss why ${\cal F}(\Psi)$ should be 
considered as a building block for ${K}(\Psi,\bar \Psi)$. 
First of all, one can 
observe that the geometry determined by ${\cal F}$ is that of the
Riemann sphere with three punctures. 
Then, by $S$--duality, modular  invariance and general 
arguments, it is quite natural to believe that $ K$ should be 
a well--defined function on ${\cal M}_{SU(2)}$, 
that is a real ``function'' of 
$u,\bar u$. On the other hand the inversion formula (\ref{f.2})
tells us that we can express $u$ by means of ${\cal F}(\Psi)$. Therefore,
${\cal F}(\Psi)$ is the building block for ${K}(\Psi,\bar \Psi)$. This is a 
useful result since, as we will see, it implies  a differential equation for 
${K}(\Psi,\bar \Psi)$,  which is the non--chiral analogue of 
(\ref{f.2}). On the other hand 
(\ref{f.2}),
which is equivalent to a second--order equation, is actually a 
(anomalous) superconformal Ward identity \cite{HW}. Then, the equation 
we will get should be interpreted as a non--chiral 
superconformal Ward identity.

The request of modular invariance indicates that 
$K$ should be constructed in terms of the geometrical building blocks of 
the thrice--punctured Riemann sphere ${\cal M}_{SU(2)}$.
The comparison between  the asymptotics (\ref{bhv2}) and (\ref{hasthestr}) 
suggests that the Poincar\'e metric should have a r\^{o}le in defining 
$K$. 
In particular, we observe that, in order to be well--defined on 
the $u$--moduli space, the logarithmic terms should come out of a
function which has to be globally defined. This would also respect
the symmetries of the theory.
The above analysis suggested the following proposal \cite{MATK}: 
\begin{equation}
\label{marco}
{K}(\Psi,\bar \Psi)=\alpha{e^{-\varphi({\cal G}(\Psi),
\overline{{\cal G}(\Psi)})}
\over |{\cal G}^2(\Psi) - \Lambda^4|}
\ \ , 
\label{ganzo}
\end{equation}
where $\alpha$ is a real constant to be determined via an explicit 
calculation, and 
\begin{equation}
e^{\varphi(u,\bar u)}={|\partial_u\tau|^2\over ({\Im}\,\tau)^2}
\ \ ,
\label{poAj}
\end{equation}
is the Poincar\'e metric on ${\cal M}_{SU(2)}$.
The expression (\ref{ganzo}) 
can also be written in the form 
\beq 
K(\Psi,\bar \Psi)=4\alpha\pi^2
{e^{2\varphi_{SW}}|{\cal G}^2(\Psi) - \Lambda^4|}
\ \ ,
\eeq 
or
\begin{equation}
{K}(\Psi,\bar \Psi)=
2\alpha\pi e^{\varphi_{SW}({\cal G}(\Psi),\overline{{\cal 
G}(\Psi)})-{\varphi\over 2}({\cal G}(\Psi),\overline{{\cal G}(\Psi)})}
\ \ ,
\label{ganzotris}
\end{equation}
where
\begin{equation}
e^{\varphi_{SW}(u,\bar u)}=|\partial_u a|^2{\Im}\, \tau
\ \ ,
\label{Pspos}
\end{equation}
is the Seiberg--Witten metric on ${\cal M}_{SU(2)}$.

Let us now consider the geometrical meaning of $ K(\Psi,\bar\Psi)$.
According to (\ref{ganzotris}) 
the $(1/2,1/2)$--differential $K$ is proportional to
the Seiberg--Witten metric times
the inverse of the
square root of the  Poincar\'e metric.
The interesting point is that the structure
of (\ref{Pspos})  does not prevent us from  considering 
for $K$ a suitable modification of the Liouville equation which 
is satisfied by the Poincar\'e metric. In particular, 
looking at the structure of 
(\ref{Pspos}),  it is easy to see that after a sufficient number 
of times one acts with the derivative operators,
the effect of the Seiberg--Witten metric on the Liouville 
equation can be eliminated.
In particular, setting
\begin{equation}
 Y(a,\bar a) =K(a,\bar a)\partial_{a}\partial_{\bar a}\ln K(a,\bar a)
\ \ ,
\label{ojw}
\end{equation}
we have the ``non--chiral superconformal Ward 
identity''\footnote{We thank Gaetano Bertoldi for interesting
discussions on this equation.}
\begin{equation}
\partial_{\bar a} \partial_a \ln Y(a,\bar a)=0
\ \ .
\label{oidjhg}
\end{equation}

\subsection{\boldmath ${K(\Psi,\bar \Psi)}$ from 
the \boldmath ${\beta}$--function}
\noindent
In \cite{BoMa} the renormalisation group equation (RGE) 
and the exact $\beta$--function 
was derived in the $SU(2)$ case. Also, similar structures
have been considered in the framework of the 
Witten--Dijkgraaf--Verlinde--Verlinde equation in
the $SU(3)$ case \cite{BoMa2}. It would be interesting to
understand the scaling properties of $K$. 
As it is constructed in terms of ${\cal F}$, one could  
imagine that the RGE for ${\cal F}$
should play a role. The RGE, derived in \cite{BoMa}, is
\begin{equation}
\partial_\Lambda{\cal F}(a,\Lambda)=
{\Lambda\over \Lambda_0}\partial_{\Lambda_0}{\cal F}(a_0,\Lambda_0)
e^{-2\int_{\tau_0}^\tau {dx
\beta^{-1}(x)}}\ \ ,
\label{renormalgroup3bis}
\end{equation}
where 
\begin{equation}
\beta(\tau)=
\Lambda \left(\partial_\Lambda\tau\right)_u\ \ ,
\label{beta}
\end{equation}
is the $\beta$--function.
Remarkably, the $\beta$--function admits a geometrical interpretation 
as the chiral block for the Poincar\'e metric, namely \cite{BoMa}
\begin{equation}
ds_P^2= \left|{\beta\over 2u\ {\Im}\, \tau}\right|^2|du|^2
=e^{\varphi}|du|^2\ \ .
\label{BetaLiouville}
\end{equation}
On physical grounds it is clear that,
the $\beta$--function should vanish at $u=0$.
However, this degeneracy 
should not appear in the relevant geometrical objects. Remarkably, 
this is actually the case. To be more precise, 
the above expression for $K$ admits the equivalent
general representation
\begin{equation}
K(a,\bar a)=4\alpha \pi {|{\cal G}(a)|({\Im}\, \tau)^2\over 
|\beta||\partial_a {\cal G}(a)|^2}\ \ .
\label{azzarolina}
\end{equation}

\subsection{The \boldmath${1\leq N_F\leq 4}$ Case}
\noindent
As the above expression for $K$ 
does not refer to a particular underlying geometry, we can 
consider (\ref{azzarolina}) as a general model--independent expression for 
$K$. In particular, observe that its 
asymptotic expansion can be performed by just using the one for 
the prepotential ${\cal F}$. 
However, there is still another equivalent form for $K$ which is particularly
useful in order to perform asymptotic analyses. We 
have in mind the fact that, 
in the presence of massless hypermultiplets, 
only instantons with even $k$  contribute.
Then,  in order to get a suitable expression for $K$,  we introduce
the function \cite{BoMa}
\begin{equation}
\beta^{(a)}(\tau)=
\Lambda \left(\partial_\Lambda\tau\right)_a
\ \ ,
\label{betas}
\end{equation}
whose relation with the $\beta$--function is \cite{BoMa}
\begin{equation}
\beta(\tau)=
2u{\partial_ua\over a}\beta^{(a)}(\tau)
\ \ .
\label{beta3}
\end{equation}
By (\ref{azzarolina}) and (\ref{beta3}) we have
\begin{equation}
K(a,\bar a)=2\alpha \pi {|a|(\Im\, \tau)^2\over 
|\beta^{(a)}||\partial_a {\cal G}(a)|}\ \ .
\label{azzarolinabisse}
\end{equation}

To better illustrate the r\^{o}le of the $\b$--function in the 
non--holomorphic contribution, we use a result in \cite{BoMa} where it was 
shown that 
\begin{equation}
u=\Lambda^2e^{-2\int_{\tau_0}^\tau dx \beta^{-1}(x)}\ \ ,
\label{aaaqdoi}
\end{equation}
where $u(\tau_0)=\Lambda^2$ (in the $N_F=0$ case, $\tau_0=0$). 
Then $K$ has the form
\begin{equation}
K(a,\bar a)= \alpha\pi \left|{a\over\Lambda}\right|^2|F|^2
e^{\int_{\tau_0}^\tau \beta^{-1}+
\overline{\int_{\tau_0}^\tau \beta^{-1}}}\left(\Im\, \tau\right)^2\ \ ,
\label{dsoqijd}
\end{equation}
where 
\begin{equation}
F(a,\bar a)={\beta^{1/2}\over \beta^{(a)}}\ \ .
\label{qwiodj}
\end{equation}
Thanks to (\ref{aaaqdoi}) and (\ref{dsoqijd}) it follows that 
$K$ satisfies the RGE 
\begin{equation}
\Lambda\left({\partial_\Lambda K(a,\bar a)}\right)_{a,\bar 
a}=
2\left[
\mbox{R}\mbox{e}\, \left({\beta^{(a)}\over 
\beta}+\beta^{(a)}\partial_\tau\ln F\right)+
{\Im\, \beta^{(a)}\over \Im\, \tau} -1\right]K(a,\bar a)\ \ .
\label{spo1ws}
\end{equation}

One can check that
when only instantons with even $k$ contribute
to ${\cal F}$, then this would also  be the case for 
the expression (\ref{azzarolinabisse}) for $K$.

Finally we note that in the $N_F=4$ case the above construction 
breaks down.
In particular,
in this case the underlying geometry is trivial. As a consequence, 
the non--trivial global aspects of moduli spaces, which actually 
generate 
non--perturbative corrections, do not arise for $N_F=4$. This is already
clear for the chiral part ${\cal F}$ which is proportional
to $a^2$. As in general the function $K$ is built in terms
of ${\cal F}$, we see that there is no way to get non--holomorphic 
contributions to $K$ but the one--loop term, whose 
structure has a global meaning since the underlying geometry
is trivial.

\section{Non--perturbative contributions to \boldmath ${K(\Psi, \bar{\Psi})}$}
\setcounter{equation}{0}
Let us now discuss the series expansion for $K(\Psi,\bar\Psi)$ in the case 
of SYM theory. 
We can rewrite (\ref{marco}) as
\beq\label{kappa}
K(\Psi,\bar{\Psi})=\frac{64\al}{\pi^2}\frac{|{\cal G}^2(\Psi)
-4\L^4|(\Im\, \t(\Psi))^2}
{|\Psi|^4 |\hat{\t}(\Psi)-\t(\Psi)|^4}
\ \ ,
\eeq
where 
\beq
\hat{\t}(\Psi)=\frac{1}{\Psi}\frac{\p \cf}{\p \Psi}
\ \ .
\eeq 
It can be fixed by using the result in \cite{ketov};
however, this is not enough to get a complete check of the validity of
(\ref{azzarolina}) and (\ref{azzarolinabisse}),  as we will discuss
in the following.

Expanding (\ref{kappa}) up to the order relevant  to 2--instanton 
calculations, and neglecting purely chiral or antichiral
terms, we find
\beqa
K(x,\bar{x})&\simeq&\al\biggl\{
\ln x\ln\bar{x}+x^4 (3\ln\bar{x}-
2\ln^2 \bar{x}-2\ln\bar{x}\ln x)+
\nonumber\\
& &\left.x^8 \left(-\frac{21}{2}
\ln x\ln\bar{x}-\frac{21}{2}\ln^2 \bar{x}+\frac{57}{8}
\ln\bar{x}\right)+\right.\\
& &x^4 \bar{x}^4 \left(\frac{9}{4}-6\ln\bar{x}
+2\ln^2 \bar{x}+4\ln x\ln\bar{x}\right)+{\mbox{\rm h.}}\mbox{\rm c.}
\biggr\}
\nonumber
\ \ ,
\eeqa
where $x=\L/\Psi$.

Let us briefly comment on the functional dependence of the various terms 
appearing in the expansion.
The first logarithmic term represents the one--loop 
perturbative contribution 
to $K(\Psi,\bar{\Psi})$ which was first derived in \cite{DEWIT}; 
it is to be noted that 
there are no higher--order (higher--loop) 
logarithmic corrections to $K(\Psi,\bar{\Psi})$.
This seems to be confirmed
by recent results found in 
\cite{ZANON},  where the two--loop correction to the 
effective Lagrangian (\ref{lagn=2n=2})  is shown to vanish. 
As far as the terms $x^{4k}\ln\bar{x}$ are concerned, they appear 
explicitly in the $k$--instanton calculations, while the terms 
with  $x^{4k}\ln\bar{x}\ln x$ and  $x^{4k}\ln^2\bar{x}$  are 
expected to be one--loop corrections around the $k$--instanton configuration.
As a matter of fact, in this case there are no constraints coming 
{}from holomorphicity requirements and from the anomalous $U(1)_{R}$ symmetry 
which forbid the existence of loop corrections around instanton 
configurations \cite{sei}. 
Finally, the terms  $x^{4m}\bar{x}^{4n}$ and logarithmic 
corrections are expected to represent $m$--instanton/$n$--antiinstanton 
contributions and loop corrections around this configuration.  
In the case of SQCD this situation is simply modified in the presence of 
massless hypermultiplets, since  the expansion contains only 
non--perturbative contributions from m--instanton/n--antiinstanton 
where $m,n$ are even numbers and one--loop corrections around these 
configurations. 
In the sequel we will perform 1-- and 2--instanton calculations which 
will give contributions to $K(\Psi,\bar{\Psi})$ of the form expected from 
the conjecture in  \cite{MATK}.
Let us now make a remark which will become clear after the instanton 
computation will be performed.
If we differentiate $K( x, \bar{x} )$ twice with respect to $x$ and twice
with respect to $\bar{x}$  (to obtain $K_{x x \bar{x}\bar{x} }$),
the terms containing the $\ln x \bar{x}$  and the 
$\ln^2  x \bar{x}$ give contributions which sum.
Therefore, for an unambiguous check of the  conjectures
(\ref{azzarolina}),   (\ref{azzarolinabisse}),  
one needs not only 1--instanton or 2--instanton but also mixed 
1--instanton/1-antiinstanton results and perturbative corrections 
around all the aforementioned configurations.
Anyway, as a first step towards the check these proposals,
we now compute the non--perturbative (1--instanton and 2--instanton) 
contributions to $K(\Psi , \bar{\Psi} )$.

In terms of $N=1$ superspace the four--derivative term reads \cite{HENN}:
\beqa
& &\frac{1}{16}\int d^2 \th d^2 \bar{\th} \Bigl[ K_{\phi \bar{\phi}} 
(\Phi,
\bar{\Phi}) \bigl( D^\al D_\al \Phi \bar{D}_{\dot{\al}}
\bar{D}^{\dot{\al}} \bar{\Phi} + 2 \bar{D}_{\dot{\al}} D^\al \Phi
D_\al \bar{D}^{\dot{\al}} \bar{\Phi} + 4 D^\al W_\al
\bar{D}_{\dot{\al}} \bar{W}^{\dot{\al}} - 
\nonumber \\
& & 4 D^{(\al} W^{\b)}
D_{(\al} W_{\b)} - 4  \bar{D}_{(\dot \al} \bar{W}_{\dot \b)}
\bar{D}^{(\dot \al}  \bar{W}^{\dot \b)} 
 - 2  D^\al D_\al (W^\b
W_\b) - 2 \bar{D}_{\dot{\al}} \bar{D}^{\dot{\al}} (
\bar{W}_{\dot{\b}} \bar{W}^{\dot{\b}} ) \bigr)  -
\nonumber \\
& & 2 K_{\phi \phi \bar{\phi}} (\Phi, \bar{\Phi}) W^\al W_\al
D^\b D_\b \Phi - 2 K_{\phi \bar{\phi} \bar{\phi}} (\Phi, \bar{\Phi})
\bar{W}_{\dot{\al}} \bar{W}^{\dot{\al}} \bar{D}_{\dot{\b}} 
\bar{D}^{\dot{\b}} \bar{\Phi} +
\nonumber \\
& &  K_{\phi \phi \bar{\phi} \bar{\phi}} (\Phi, \bar{\Phi}) \left( - 8 W^\al
D_\al \Phi \bar{W}_{\dot{\al}} \bar{D}^{\dot{\al}} \bar{\Phi} + 4
W^\al W_\al \bar{W}_{\dot{\al}} \bar{W}^{\dot{\al}} \right) \Bigr]
\ \ ,
\eeqa
where $K_{\phi}=\p K/\p \ph$.
When written in the $x$--space this Lagrangian contains a 
four--field strength vertex which is the one we will focus our 
attention on in our calculations:
\beqa
& &\frac{1}{4}\int d^4 x d^2 \th d^2 \bar{\th} \ 
K_{\phi \phi \bar{\phi} \bar{\phi}} (\Phi, \bar{\Phi}) W^\al W_\al 
\bar{W}_{\dot{\al}} \bar{W}^{\dot{\al}}=
\nonumber \\
& &\frac{1}{256}K_{a a \bar{a} \bar{a}}(a,\bar{a})\int d^4 x \ \Tr (\s^{ab}
\s^{cd}) \Tr(\sb^{ef}\sb^{gh})F_{ab}F_{cd}F_{ef}F_{gh} \ .
\eeqa
Thus, the correlator we intend to study is 
\beq
\label{cicciop}
\langle F_{\m\n}(x_{1})F_{\r\s}(x_{2})F_{\l\t}(x_{3})F_{\k\th}(x_{4})
\rangle
\ \ .
\eeq
\subsection{The \boldmath ${k=1}$ Semiclassical Computation}
\noindent
The relevant configuration which contributes to this 
Green function is dictated by 
the sweeping--out procedure at the next--to--leading--order of \cite{dkm}:
\beqa\label{fmunu}
F_{\m\n}=F_{\m\n}^{(0)}+i\xi(x)\s_{[\n}D_{\m]}\lb^{(0)}+
2i\lb^{(0)}\sb_{\m\n}\bar{\eps}+2ig\xi^2(x) \lb^{(0)}\sb_{\m\n}\lb^{(0)}
\ \ ,
\eeqa
where 
$F_{\m\n}^{(0)}$ satisfies the equation
\beq
D^{\m}F_{\m\n}^{(0)}=-2ig [\ph^{\dagger}_{\rm cl},D_{\n}\ph_{\rm cl}] 
\ \ ,
\eeq
with 
\beq
\ph_{\rm cl}=\frac{x^2}{x^{2}+\r^2}a^{c}(\s^c /2)
\ \ ,
\eeq
and
\beq
\xi(x)=\xi+\r^{-1}x^{\m}\s_{\m}\bar{\varepsilon}\ \ ,
\eeq 
\beq
\lb^{(0)}=-i\sqrt{2}\xi^{\prime}\Ds \phi^{\dagger}_{\rm cl}\ \ .
\eeq  
Here,   for simplicity,  $x$  stands  for $x-x_0$, where 
$x_0$ is the centre
of the 1--instanton configuration, and $\rho$ is its size; 
finally 
$\bar{\eta}^{\prime a}_{\m\n}=R_{ab}\bar{\eta}^{a}_{\m\n}$, where 
$R_{ab}$ is an $SU(2)$ rotation matrix which corresponds to global colour 
rotations. 

We start by rederiving the  result of \cite{YUNG} 
for the 1--instanton case 
in a different way.
In the case $k=1$, the $N=2$ SYM measure on the moduli space is 
simply  \cite{th,ber}
\beq
\label{meask=1}
\int d^3 \Theta d^4 x_{0}\frac{\de\r}{\r^5} 
\frac{2^7 \pi^6}{g^{8}} (\m\r)^8 
\left(\frac{16 \pi^2 \mu}{g^2}\right)^{-2}d^2\xi d^2\xi^{\prime}
\left(\frac{32\pi^{2}\r^2 \m}{g^2}\right)^{-2}d^2\bar{\eps}
d^{2}\bar{\eps}^{\prime}\exp(-S_{inst})
\ \ ,
\eeq
where $S_{inst}$ is the sum of the classical action, the Higgs and the 
Yukawa terms,
and $\Theta^a$, $a=1,2,3$ denotes the moduli associated with 
global colour rotations.

\noindent
We observe first that $F_{\m\n}$ does not contain the superconformal 
collective coordinate $\bar{\eps}^{\prime}$ so that the integration 
over the superconformal fermionic coordinates 
must be completely saturated by 
the Yukawa action and we can ignore the terms in $F_{\m\n}$ which depend
on the fermionic coordinate $\bar{\eps}$. Therefore in evaluating the 
correlator  (\ref{cicciop}), only the first, the second and the fourth
term in the r.h.s. of (\ref{fmunu}) will be of interest.
To lowest order in $g^2 \r^2 |a|^2$, 
$F_{\m\n}^{(0)}$ 
becomes 
\beq\label{superinst}
F_{\m\n}^{\rm cl}=\frac{4 \r^2}{g}\frac{1}{x^2 (x^2+\r^2)^2}(-x^2
\bar{\eta}^{\prime a}_{\m\n}+2x_{\l}x_{\n}\bar{\eta}^{\prime a}_{\m\l}+
2x_{\l}x_{\m}\bar{\eta}^{\prime a}_{\l\n})\frac{\s^{a}}{2} 
\ \ ,
\eeq
and the term proportional to 
$\xi^2$ is negligible. 
Then, 
in order to saturate 
the integration over the supersymmetric collective coordinates 
$\xi, \xi^{\prime}$, 
the product in (\ref{cicciop}) boils down to 
\beq
F_{\m\n}^{\rm cl}(x_{1})F_{\r\s}^{\rm cl}(x_{2}) 
[ i\xi\s_{[\n}D_{\m]}\lb^{(0)} (x_3)]
[i\xi\s_{[\n}D_{\m]}\lb^{(0)} (x_4)]
\ \ .
\eeq
Now we have to extrapolate the relevant long--distance effective $U(1)$ 
fields (\ref{superinst}):
\beq
F^{(3) {\rm cl, LD}}_{\m\n}(x)=\frac{4 \r^2}{g}\cdot{1\over x^6}
(-x^2
\bar{\eta}^{\prime 3}_{\m\n}+2x_{\l}x_{\n}\bar{\eta}^{\prime 3}_{\m\l}+
2x_{\l}x_{\m}\bar{\eta}^{\prime 3}_{\l\n})
\ \ ,
\eeq
and 
\beq\label{bilin}
i\xi\s_{[\n}\p_{\m]}\lb^{(0)}_{LD}
\ \ ,
\eeq
where $\lb^{(0)}_{LD}=-i\sqrt{2}\xi^{\prime}\ \ds 
\ph^{\dagger LD}_{\rm cl}$ and the suffix LD stands for long--distance. 
In this limit the covariant derivative becomes 
a simple one.
In \cite{dkm} a nice 
relationship between the scalar Higgs field and the Higgs action
in the long--distance limit was derived, 
\beq
\ph^{\dagger}_{\rm cl, LD}=\bar{a}-a^{-1}S_{H}G(x,x_{0})
\ \ ,
\eeq
where 
$G(x,x_{0})=1/4\pi^2 (x-x_{0})^2$ is the massless scalar propagator.
As a consequence of this observation 
it is possible to recast (\ref{bilin}) into the 
form 
\beq
\frac{\sqrt{2}}{2}\frac{\p}{\p a}S_{H}\xi\s^{ab}
\xi^{\prime}G_{\m\n,ab}(x,x_{0}) 
\ \ ,
\eeq
where $G_{\m\n,ab}(x,x_{0})$ is the gauge--invariant propagator of the 
$U(1)$ field strength
\beq
G_{\m\n,ab}(x,x_{0})=(\d_{\n b}\p_{\m}\p_{a}-\d_{\n a}\p_{\m}\p_{b}-
\d_{\m b}\p_{\n}\p_{a}+\d_{\m a}\p_{\n}\p_{b})G(x,x_{0})
\ \ .
\eeq
The integration on the superconformal collective coordinates, which are lifted 
in the background of the constrained instanton, is completely saturated 
by the Yukawa action $S_{Y}$,  and one gets
\cite{nsvz2}
\beq
\int d^2\bar{\eps}
d^{2}\bar{\eps}^{\prime}\exp(-S_{Y})=-2^{9}\pi^{4}g^{-2}\r^{4}
\bar{a}^{2}
\ \ .
\eeq
The key observation is that the only dependence on the coordinates $\Theta$ is 
due to the insertion of  $F_{\m\n}^{(3){\rm cl}}$ and that, in the 
long--distance limit,
\beqa
\int_{SU(2)/Z_{2}}d^3 \Theta 
F_{\m\n}^{(3){\rm cl}}(x_{1})F_{\r\s}^{(3){\rm cl}}(x_{2})&=&
\frac{8\pi^2}{3}F_{\m\n}^{a\, {\rm cl}}(x_{1})F_{\r\s}^{a\, {\rm cl}}(x_{2})=\\
&=&-\frac{16\pi^6 \r^4}{3g^2}\Tr(\sb^{ef}\sb^{gh})G_{\m\n,ef}(x_{1},x_{0})
G_{\r\s,gh}(x_{2},x_{0})
\ .
\nonumber
\eeqa
Taking into account the other two insertions which saturate the 
integration over $\xi,\xi^{\prime}$ we finally obtain
\beqa
& &\langle F_{\m\n}(x_{1})F_{\r\s}(x_{2})F_{\l\t}(x_{3})F_{\k\th}(x_{4})
\rangle_{k=1}=-\frac{15}{64\pi^{2}}\frac{\L^{4}}{g^4 \bar{a}^2 a^{6}}
\int d^{4}x_{0}\Tr(\s^{ab}\s^{cd})\nonumber\\ 
& &G_{\m\n,ab}(x_{1},x_{0})
G_{\r\s,cd}(x_{2},x_{0})\Tr(\sb^{ef}\sb^{gh})G_{\l\t,ef}(x_{3},x_{0})
G_{\k\th,gh}(x_{4},x_{0})\ \ . 
\eeqa
On the other hand the computation of the four--field strength vertex 
making use of the effective Lagrangian yelds
\beqa
& &\langle F_{\m\n}(x_{1})F_{\r\s}(x_{2})F_{\l\t}(x_{3})F_{\k\th}(x_{4})
\rangle_{L-eff}=\frac{3}{32}K_{a a \bar{a} \bar{a}}(a,\bar{a})\int 
d^4 x \Tr(\s^{ab}\s^{cd})
\nonumber \\
& & G_{\m\n,ab}(x_{1},x_{0})
G_{\r\s,cd}(x_{2},x_{0})\Tr(\sb^{ef}\sb^{gh})G_{\l\t,ef}(x_{3},x_{0})
G_{\k\th,gh}(x_{4},x_{0})\ \ ,
\eeqa
which finally reproduces the result \cite{YUNG}
\beq
K(a,\bar{a})=\frac{1}{8\pi^{2}g^{4}}\frac{\L^{4}}{a^{4}}\ln\bar{a}
\ \ .
\eeq
We can rewrite the 1--instanton correlator in a form which is well--suited to 
the generalisation to SQCD with 
$1 \leq N_{F}\le 4$ 
massive hypermultiplets in the fundamental representation 
of $SU(2)$ (in the case in which 
at least one hypermultiplet is massless 
the non--perturbative contributions are expected to 
come only from $m$--instanton 
$n$--antiinstanton configurations where $m$, $n$ are even),
\beqa\label{1inst}
& &\langle F_{\m\n}(x_{1})F_{\r\s}(x_{2})F_{\l\t}(x_{3})F_{\k\th}(x_{4})
\rangle_{k=1}=\frac{\pi^4}{2}\left(\int d^{4}x_{0}\Tr(\s^{ab}\s^{cd})
G_{\m\n,ab}(x_{1},x_{0})\right.
\nonumber \\
& &\left. G_{\r\s,cd}(x_{2},x_{0})
\Tr(\sb^{ef}\sb^{gh})G_{\l\t,ef}(x_{3},x_{0})
G_{\k\th,gh}(x_{4},x_{0})\right)\frac{\p^{2}}{\p a^{2}}\left[
\int\de\tilde{\m}_1\r^4\right]
\ \ ,
\eeqa
where $d\tilde{\m}_1$ is the ``reduced" instanton measure obtained by 
extracting from the full measure the integration over the bosonic and 
fermionic translational coordinates \cite{dkm,GAB}.
This formula generalises immediately by exchanging $\de\tilde{\m}_1$ 
with $\de\tilde{\m}^{N_{F}}_1$ \cite{dkm},  where
\beq
\int \de\tilde{\m}^{N_{F}}_1=-\frac{1}{16\pi^2 g^4}\frac{\L^{4-N_{F}}_{N_{F}}}
{a^2}\prod_{i=1}^{N_{F}}m_{i}\ \ ,
\eeq
and $m_{i}$ is the mass of the $i$--th hypermultiplet. 
By doing this we obtain 
\beq
K(a,\bar{a})|_{N_{F}}=\frac{1}{8\pi^{2}g^4}\frac{\L^{4-N_{F}}_{N_{F}}}
{a^2}\ln\bar{a}\prod_{i=1}^{N_{F}}m_{i}
\ \ ,
\eeq
which is in complete agreement with one of the results obtained  in 
\cite{YUNGQCD}. 
It is to be noted that in the case $N_{F}=4$ the $\b$-function 
vanishes identically so that the scale $\L_{N_{F}}$ must be replaced by 
$q=\exp(2 i\pi\t_{\rm cl})$,  where
$\t_{\rm cl}$ is defined in  (\ref{ol.4}).
\subsection{The \boldmath ${k=2}$ Computation}
Let us now describe the calculation of the 2--instanton contribution 
to the real function $K(\Psi , \bar{\Psi} )$. 
Again, the Green function which we 
are going  to study is the simplest one, the four--field strength one. We 
will then be able to immediately generalise  our calculation 
to the case of SQCD 
and to check the validity of the 
non--renormalisation theorem in the case $N_{F}=4$ 
found in \cite{SD}.

We start by briefly recalling  how to determine gauge field 
configurations for a generic winding number $k$.
The instanton field can be conveniently written in terms 
of the Atiyah--Drinfeld--Hitchin--Manin
(ADHM)  construction \cite{ADHM,aty}. 
To find an instanton  solution  of winding number $k$, one
introduces a $(k+1)\times k$ quaternionic matrix
\beq
\Delta=a+bx \ \ ,
\label{f.4}
\eeq
where 
$x$ denotes a point  of the one--dimensional 
quaternionic space $\bb{H}\equiv \C^{2}\equiv\R4$,
$x=x^\mu \sigma_\mu$.
\footnote{We use the conventions of \cite{GAB}.}
The gauge connection is then 
written in the form 
\beq
A^{cl}_\mu=U^{\dagger}\p_\mu U\ \ ,
\label{trecinque}
\eeq
where $U$ is a $(k+1)\times 1$ matrix of quaternions providing an
orthonormal frame of $\Ker \Delta^\dagger$, \ie\ 
\beq
\Delta^\dagger U = 0\ \ ,
\label{f.5}
\eeq
\beq
U^\dagger U =\uno_2
\ \ .
\label{f.6}
\eeq
The constraint (\ref{f.6}) ensures that $A_\mu^{cl}$ 
is an element  of the Lie algebra of the $SU(2)$ gauge group.
The condition of self--duality on the field strength of 
(\ref{trecinque}) is imposed by restricting the matrix $\Delta$ 
to obey
\beq
\Delta^\dagger\Delta=f^{-1}\otimes\uno_2\ \ ,
\label{realita}
\eeq 
with $f$ an invertible hermitian $k\times k$ matrix 
(of real numbers). 
The reparametrisation invariances of the ADHM construction
\cite{osb}
can be used to simplify the expressions
of $a$ and $b$. Exploiting this fact, in the
following we will 
choose the matrix $b$ to be
\beq
b=-\pmatrix {0_{1\times k}\cr\uno_{k\times k}}
\ \ .
\label{boh}
\eeq
{}From (\ref{trecinque}), one can compute the field strength of the gauge 
field, which reads
\beq
F_{\m\n}^{\rm cl}=2U^\dagger b\s_{\m\n}f b^{\dagger} U
\label{bohboh}
\ \ .
\eeq
In the so--called singular gauge, one has 
\beqa
U_{0}&=&\s_{0}\left(1-\half f_{lm}\tr
v_{l}\vb_{m}\right)^{1/2}
\ \ ,
\nonumber \\
U_{p}&=&-\frac{1}{|U_{0}|^2}\Delta_{p l}f_{lm}\vb_{m}U_{0}
\ \ ,
\eeqa
where  $v_{p}=\Delta_{0p}$  
and $l,m,p=1, \ldots, k$. In the following  we will need only
the long--distance limit of these functions, 
\beqa
& &\Delta_{p l}\sim b_{p l}x\ \ ,
\qquad f_{lm}\sim\frac{1}{x^2}\d_{lm}
\ \ ,
\nonumber\\
& &U_{k}\sim-\frac{1}{x^2}x\vb_{k}U_{0}\ \ ,
\qquad U_{0}\sim\s_{0}
\ \ ,
\\
& &\Delta_{0l}\sim 0\nonumber
\ \ .
\eeqa
When $k=2$ the most general instanton configuration  can be  
written starting from the ADHM matrix 
\beqa
\label{ffc}
a=\left(\begin{array}{cc}v_{1}&v_{2}\\
x_{0}+e&d\\
d&x_{0}-e\end{array}\right)
\ \ .
\eeqa
Here
\beq
d=\frac{e}{4|e|^2}(\bar{v}_{2}v_{1}-\bar{v}_{1}v_{2})
\ \ ,
\eeq
as a consequence of the ADHM defining equations
\cite{csw}.

\noindent
The fermionic zero--modes 
$
\lambda_{\beta \dot{A}}^{(0)}
$
are easily deduced from 
the gauge field zero--modes \cite{osb}
\beq
Z_\mu=U^\dagger C\bar\sigma_\mu f b^\dagger U-
U^\dagger bf\sigma_\mu C^\dagger U
\ \ ,
\label{f.9}
\eeq
by recalling  that, 
due to $N=2$ SUSY,
\beq
\lambda_{\beta \dot{A}}^{(0)} = 
\sigma^{\mu}_{\beta \dot{A}} Z_{\mu} 
\ \ ,
\label{dual}
\eeq
($\dot{A}=1,2$ labels the two SUSY charges and $\beta=1,2$ is a spin 
index).
For (\ref{f.9}) to be transverse zero--modes, the $(k+1)\times k$ 
matrix $C$ (for a generic instanton number $k$)
must satisfy
\beq
\Delta^\dagger C=(\Delta^\dagger C)^T
\ \ ,
\label{f.10}
\eeq
where the superscript $T$ stands for transposition of the
quaternionic
elements of the matrix (without transposing the quaternions 
themselves).
The number of $C$'s 
satisfying (\ref{f.10}) is $8k$ \cite{osb}. 
We also need the form of the matrix $C$ appearing  in (\ref{dual}), 
which is constrained by (\ref{f.10}) to describe the 
zero--modes of the $N=2$ gauginos $\lambda_{\beta \dot{A}}^{(0)}$.
To parallel the form of (\ref{ffc}), 
we shall put   
\beqa
C_{\dot{1}}=
\left(\begin{array}{cc}
\m_{1}&\m_{2}\\
4\xi+\eta&\d\\
\d&4\xi-\eta\end{array}\right)
\ \ ,
\eeqa
\beqa
C_{\dot{2}}=\left(\begin{array}{cc}
\n_{1}&\n_{2}\\
4\xi^{\prime}+\eta^{\prime}&\d^{\prime}\\
\d^{\prime}&4\xi^{\prime}-\eta^{\prime}\end{array}\right)
\ \ ,
\eeqa
where $\d,\d^{\prime}$ are constrained 
by (\ref{f.10}) to be
\beqa
\d&=&\frac{e}{2|e|^2}(2\bar{d}\eta+\bar{v}_{2}
\m_{1}-\bar{v}_{1}\m_{2})
\ \ ,
\\
\d^{\prime}&=&\frac{e}{2|e|^2}(2\bar{d}\eta^{\prime}+\bar{v}_{2}
\n_{1}-\bar{v}_{1}\n_{2})\nonumber
\ \ .
\eeqa
In the long--distance limit,  the 2--instanton field strength
factorises in 
\beqa
F_{\m\n}^{{\rm cl}\,LD}&=&\frac{2}{x^6}
[v_{1}\bar{x}\s_{\m\n}x\bar{v}_{1}+(v_{1}\rightarrow
v_{2})]=\\
& &\frac{1}{x^6}[v_{1}(-x^{2}\sb_{\m\n}+2x^{\r}x_{\m}\sb_{\r\n}+
2x^{\r}x_{\n}\sb_{\m\r})\bar{v}_{1}+(v_{1}\rightarrow v_{2})]\nonumber
\ \ .
\eeqa
On the other hand in \cite{dkm} it was proved that, thanks to 
the geometrical properties of 
the ADHM construction, the relationship between the 
$\xi,\xi^{\prime}$ bilinear part in (\ref{fmunu})  
and the Higgs action continues to hold for every winding number. 

We start with the $k=2$ $N=2$ supersymmetric measure,  which reads
\beqa
\frac{1}{{\cal S}_2}\int d^4 x_{0}d^4 e d^4 v_{1}
d^4 v_{2}d^2 \xi d^2 \xi^{\prime} d^2 \eta
d^2 \eta^{\prime} d^2 \m_{1} d^2 \m_{2} d^2 \n_{1}d^2 \n_{2}
\exp(-S_{inst})\left(\frac{J_{B}}{J_{F}}\right)^{1/2}
\ \ .
\eeqa
${\cal S}_{2}$ is the $k=2$ symmetry factor which eliminates 
all the redundant copies of each field configuration which appears 
in the ADHM formalism \cite{osb,dkm},
and $J_{B} (J_{F})$ is the Jacobian of the
change of variables for the bosonic (fermionic) degrees of freedom.
As in the calculation of the 2--instanton contribution to
the $N=2$ prepotential \cite{dkm}, 
we find it convenient to define the four 
combinations of the bosonic parameters:
\beqa
L&=&|v_{1}|^2+|v_{2}|^2
\ \ ,
\nonumber\\
H&=&L+4|d|^{2}+4|e|^2
\ \ ,
\nonumber\\
\O&=&v_{1}\bar{v}_{2}-v_{2}\bar{v}_{1}
\ \ ,
\\
\om&=&\frac{1}{2}\tr \ \O A_{00}
\ \ ,
\nonumber
\eeqa
where $A_{00}=\frac{i}{2}a^{c}\s^{c}$.
In terms of these new variables it is possible to write the Higgs action 
as
\beq
S_{H}=16\pi^2\left(L|A_{00}|^{2}-\frac{|\om|^2}{H}\right)=
4\pi^2 |a|^2 \left(L-\frac{|\O |^2 \cos^2 \th}{H}\right)
\ \ ,
\eeq
and the Yukawa action as
\beq
S_{{\rm Y}}=4\sqrt{2}\pi^2\Bigl[-\n_k \bar{A}_{00}\m_k+
(\bar{\om}/H)(\m_1\n_2-\n_1\m_2+2\eta\d^{\prime}-2\eta^{\prime}\d)\Bigr]
\ \ ,
\eeq
where $|\om|=\frac{1}{2}|\O||a||\cos\th|$ defines the
polar angle $\th$. Finally 
\beq
\frac{1}{{\cal S}_{2}}\left(\frac{J_{B}}{J_{F}}\right)^{1/2}\exp(-S_{\rm cl})=
2^6 \pi^{-8} \L^8 \frac{\Big| |e|^2-|d|^2\Big|}{H}\ \ .
\eeq
As in the 1--instanton case the integration over the non--supersymmetric 
fermionic coordinates is saturated by the Yukawa action which gives
\beqa
\int d^2 \eta
d^2 \eta^{\prime} d^2 \m_{1} d^2 \m_{2} d^2 \n_{1}d^2 \n_{2}\exp
(-S_{{\rm Y}})&=&-\frac{2^5 \pi^6  \bar{a}^6 \cos^2 \th}{|e|^4
H^{\prime 2}}L^2 \left[\left(1+\frac{\cos^2 \th}{H^{\prime}}\right)^2+
\right.\nonumber\\
& &\left.\frac{1-|\O^{\prime}|^{2}}{H^{\prime 2}}\sin^2 \th \cos^2 \th\right]
\ \ ,
\eeqa
where we have redefined $\O^{\prime}=\O/L, H^{\prime}=H/L$.
The integration over the variable $e$ is traded for the integration on 
$H$, {\it i.e.}
\beq
\int d^4 e \ \frac{\Big| |e|^2-|d|^2\Big|}{|e|^4}\quad\longrightarrow\quad
\frac{\pi^2}{2}\int_{L+2|\O|}^{\infty} d H \ \ .
\eeq
As far as the two insertions of $F_{\m\n}$ bilinear in $\xi,\xi^{\prime}$ are
concerned, it is possible to use a trick already exploited in the 1--instanton 
case. It  consists in writing them as a second derivative of the instanton
measure with respect to $a$ \cite{dkm}; the remaining two insertions, 
however, will have to be 
inserted and integrated explicitly. 
First of all let us 
write $v_{2}$ as a function of $v_{1},\O,L$, 
\beq
v_{2}=\left(\frac{\bar{\O}}{2}+\sqrt{|v_{1}|^2 (L-|v_{1}|^2)-
\frac{|\O|^{2}}{4}}\right)\frac{v_{1}}{|v_{1}|^2}
\ \ ,
\eeq 
and insert this form in the long--distance limit of the 2--instanton 
classical configuration. 
The integration measure over $v_{1},L,\O$ is written as
\beqa
2\int_{0}^{\infty}\de L\int_{|\O|\le L}d^3 \O
\int_{L_{-}}^{L_{+}}\de|v_{1}|^2 \frac{1}{32\sqrt
{(L_{+}-|v_{1}|^2)(|v_{1}|^{2}-L_{-})}}\int_{S^3}d^3 \Theta
\ \ ,
\eeqa
where $\int_{S^3}d^3 \Theta=2\pi^2$ is the integration over 
the global colour rotations of the first centre of the instanton 
and $L_{\pm}=\half(L\pm\sqrt{L^2-|\O|^2})$. 
On the other hand 
\beq
\int d^3 \O=L^3\int_{0}^{2\pi}\de\varphi\int_{-1}^{1}\de(\cos\th)
\int_{0}^{1}|\O^{\prime}|^2 \de|\O^{\prime}|
\ \ ,
\eeq
where $\th$ is the angle between $\O$ and the direction singled out 
by the vev of the Higgs field.
Again, as in the 1--instanton case, the key observation is that, 
in the long--distance limit, 
\beqa
& &\int d^3 \Theta F_{\m\n}^{3\, {\rm cl}}(x)F_{\r\s}^{3\, {\rm cl}}(y)=
\frac{2\pi^2}{3}F_{\m\n}^{a\, {\rm cl}}(x)F_{\r\s}^{a\, {\rm cl}}(y)=\\
& &-\frac{2\pi^6}{3}\Tr(\sb^{ab}\sb^{cd})G_{\m\n,ab}(x,x_{0})
G_{\r\s,cd}(y,x_{0})\left(L|v_{1}|^2- \frac{|\O|^2}{2}\sin^2 \th
\right)\nonumber
\ \ .
\eeqa
Putting everything together one obtains the following integral 
for the correlator:
\beqa
& &\int d^4 x_{0}\int_{0}^{1}d |\O^{\prime}||\O^{\prime}|^6 
\int_{-1}^{1}d (\cos\th)\cos^2 \th
\int_{1+2|\O^{\prime}|}^{\infty}\frac{d H^{\prime}}{H^{\prime 3}}
\int_{0}^{\infty}d L L^{7}\nonumber\\
& &[1-|\O^{\prime}|^2 \sin^2 \th]
(-4\pi^{14})\L^{8}\bar{a}^6 
\left[\left(1+\frac{\cos^2 \th}{H^{\prime}}\right)^2+
\frac{1-|\O^{\prime}|^{2}}{H^{\prime 2}}\sin^2 \th \cos^2 \th\right]
\nonumber\\
& &\Tr(\sb^{ab}\sb^{cd})
G_{\m\n,ab}(x_{1},x_{0})
G_{\r\s,cd}(x_{2},x_{0})\Tr(\s^{ef}\s^{gh})G_{\l\t,ef}(x_{3},x_{0})
G_{\k\th,gh}(x_{4},x_{0})\nonumber\\
& &\frac{\p^2}{\p a^2}\exp\left[-4\pi^2 L|a|^2 \left(1-\frac{|\O^{\prime}|^2
\cos^2 \th}{H^{\prime}}\right)\right]
\ \ ,
\eeqa
and, after a trivial integration on $L$ we get 
\beqa
& &\int d^4 x_{0}\int_{0}^{1}\de|\O^{\prime}||\O^{\prime}|^6 
\int_{-1}^{1}\de(\cos\th)\cos^2 \th
\int_{1+2|\O^{\prime}|}^{\infty}\frac{\de H^{\prime}}{H^{\prime 3}}
[1-|\O^{\prime}|^2 \sin^2 \th]\nonumber\\
& & \left(-\frac{5\cdot 3^4\cdot 7}{2^6\pi^2}\right)\frac{\L^{8}}
{\bar{a}^2 a^{10}}
\frac{\left(1+\frac{\cos^2 \th}{H^{\prime}}\right)^2+
\frac{1-|\O^{\prime}|^{2}}{H^{\prime 2}}\sin^2 \th \cos^2 \th}
{\left(1-\frac{|\O^{\prime}|^2\cos^2 \th}{H^{\prime}}\right)^8}
\Tr(\sb^{ab}\sb^{cd})G_{\m\n,ab}(x_{1},x_{0})\nonumber\\
& &G_{\r\s,cd}(x_{2},x_{0})\Tr(\s^{ef}\s^{gh})G_{\l\t,ef}(x_{3},x_{0})
G_{\k\th,gh}(x_{4},x_{0})
\ \ .
\eeqa
The remaining integrations over the adimensional 
variables $|\O^{\prime}|,\cos\th,H^{\prime}$ can be easily performed 
by using a standard algebraic manipulation routine and give 1/42. 
The final result is then,  restoring the explicit $g$ dependence, 
\beqa
& &\langle F_{\m\n}(x_{1})F_{\r\s}(x_{2})F_{\l\t}(x_{3})F_{\k\th}(x_{4})
\rangle_{k=2}=-\frac{5\cdot3^3}{2^7 \pi^2 g^8}\frac{\L^8}{\bar{a}^2 a^{10}}
\int d^4 x_{0}\Tr(\sb^{ab}\sb^{cd})\nonumber\\
& &G_{\m\n,ab}(x_{1},x_{0})
G_{\r\s,cd}(x_{2},x_{0})\Tr(\s^{ef}\s^{gh})G_{\l\t,ef}(x_{3},x_{0})
G_{\k\th,gh}(x_{4},x_{0})
\ \ .
\eeqa
Comparing this result to that of the effective Lagrangian  gives
\beq
\label{dvc}
K(a,\bar{a})|_{k=2}=\frac{5}{32\pi^2 g^8}\frac{\L^8}{a^8}\ln \bar{a}
\ \ ,
\eeq
which is our prediction for the 2--instanton contribution to the real 
function $K(\Psi , \bar{\Psi} )$ while the 2--antiinstanton 
configuration contribution to $K$ is simply the complex conjugate of 
(\ref{dvc}). 

Let us generalise our result to the case of $N_{F}\le 4$ massless 
hypermultiplets which receives the first non--perturbative contribution from 
the 2--instanton sector and verify the 
non--renormalisation theorem of \cite{SD} for $N_{F}=4$. 
As in the 1--instanton case (see (\ref{1inst})) 
it is possible to rewrite the four--field strength correlator as a double 
derivative of the ``reduced" measure with respect to $a$ 
\beqa
& &\langle F_{\m\n}(x_{1})F_{\r\s}(x_{2})F_{\l\t}(x_{3})F_{\k\th}(x_{4})
\rangle_{k=2}=\frac{\pi^4}{4}\frac{\p^2}{\p a^2}\left[\int\de\tilde{\m}_{2}
\left(|v_{1}|^2 L-\frac{|\O|^2}{2}\sin^2 \th\right)\right]\\
& &\int d^4 x_{0}\Tr(\sb^{ab}\sb^{cd})
G_{\m\n,ab}(x_{1},x_{0})G_{\r\s,cd}(x_{2},x_{0})\Tr(\s^{ef}\s^{gh})
G_{\l\t,gh}(x_{3},x_{0})G_{\k\th,gh}(x_{4},x_{0})\nonumber
\ \ ,
\eeqa
and the extension to the case $N_{F}>0$ is performed by substituting 
the ``reduced" measure $\de\tilde{\m}_{2}$ with $\de\tilde{\m}^{N_{F}}_{2}$ 
as defined in \cite{DKMMATT}:
\beqa
\int\de\tilde{\m}^{N_{F}}_{2}&=&-2^9\pi^7 \bar{a}^2 \L_{N_{F}}^{(4-N_{F})}
\int_{0}^{1}\de|\O||\O|^2\int_{-1}^{1}\de(\cos\th)
\int_{1+2|\O|}^{\infty}\frac{\de H}{H^{3}}\int_{S^{3}}d^{3}\Theta
\int_{0}^{\infty}\de L L\nonumber\\
& &\int_{L_{-}}^{L_{+}}\frac{\de|v_{1}|^2}
{\sqrt{(L_{+}-|v_{1}|^2)(|v_{1}|^2-L_{-})}}
\exp\left[-4 \pi^2 L |a|^2 \left(1-\frac{|\O|^2 \cos^2 \th}{H}\right)
\right]\cdot\nonumber\\
& &\cdot\sum_{n=0}^{N_{F}}\frac{M_{N_{F}-n}^{(N_{F})}}{\pi^{4n}}
\left.\frac{\p^{2n}G}{\p Z^{2n}}\right|_{Z=0}
\ \  .
\eeqa
We have dropped for simplicity the primes on $H,\O$; $G(Z)$ 
contains the contribution from the integration measure over 
the hypermultiplets and has the form 
\beqa
G(Z)&=&\left(\bar{\om}L+\frac{i Z}{8 \sqrt{2}}\right)^2 \left[\frac{\bar{a}^2}
{16}|\O|^2 L^2 +\frac{L}{2H}\bar{a}\bar{\om}L\left(\bar{\om}L
+\frac{i Z}{8 \sqrt{2}}\right)+\frac{1-|\O|^2\sin^2 \th}{4 H^2} \cdot
\right. 
\nonumber\\
& &\left.
\cdot \left(\bar{\om}L+\frac{i Z}{8 \sqrt{2}}\right)^2\right]
\exp\left[\frac{i\pi^2 Z}{\sqrt{2}H}|\O|a L \cos\th\right]
\ \ .
\eeqa
The $M_{N_{F}-n}^{(N_{F})}$ are a set of $SO(2N_{F})$ invariant 
polynomials in the masses $m_{n}$ of the hypermultiplets:
\beqa
M_{0}^{(N_{F})}&=&1\ \ ,
\nonumber\\
M_{1}^{(N_{F})}&=&\sum_{n=1}^{N_{F}}m_{n}^2\ \ ,
\nonumber\\
M_{2}^{(N_{F})}&=&\sum_{n<p}^{N_{F}}m_{n}^2 m_{p}^{2}\ \ ,
\\
\vdots& &\vdots\nonumber\\
M_{N_{F}}^{(N_{F})}&=&\prod_{n=1}^{N_{F}}m_{n}^2\nonumber
\ \ .
\eeqa
In the case of massless hypermultiplets, 
the only contribution to the correlator will come from 
the term with the $2N_{F}$--th derivative of $G(Z)$ and, writing the generic 
contribution to $K(\Psi , \bar{\Psi} )$ as 
\beq
\label{die.1}
K(\Psi , \bar{\Psi} )|_{N_F<4}=K_{2}^{(N_{F})}\frac{1}{\pi^2 g^{8}}
\left(\frac{\L_{N_{F}}}{\Psi}\right)^{2(4-N_{F})}\ln\bar{\Psi}
\ \ ,
\eeq
we find 
\beqa
K_{2}^{(0)}&=&\frac{5}{32}
\ \ ,
\qquad K_{2}^{(1)}=-\frac{3^3}{2^{10}}
\ \ ,
\nonumber\\
K_{2}^{(2)}&=&\frac{3}{2^{10}}
\ \ ,
\qquad K_{2}^{(3)}=-\frac{1}{2^{12}}
\ \ .
\eeqa
For the case $N_{F}=4$ we get
\beq
\label{die.2}
K(\Psi , \bar{\Psi} )|_{N_{F}=4}=\frac{q^2}{3^3 2^{11} \pi^2 g^8}
\ln{\bar{\Psi}}
\ \ ,
\eeq
which is a purely antichiral term. When integrated 
over the whole superspace it does not contribute to the effective action; this 
confirms thus the non--renormalisation theorem of \cite{SD}.
In the case in which there are 
massive hypermultiplets, (\ref{die.1}), (\ref{die.2})
generalises immediately to the formula 
\beq
K(\Psi , \bar{\Psi} )|_{N_F}=\sum_{n=0}^{N_F}
M_{N_F-n}^{(N_F)} 
K_{2}^{(N_{F})}\frac{1}{\pi^2 g^{8}}
\left(\frac{\L_{N_{F}}}{\Psi}\right)^{2(4-N_{F})}\ln\bar{\Psi}
\ \ ,
\eeq
provided that one replaces $\L_{N_F}^{2(4-N_{F})}$ with $q^2$ when $N_F=4$.
In this case the non--renormalisation theorem of \cite{SD}, as already noted 
in \cite{DKMNEW}, is spoilt by the presence of other energy scales 
represented by the masses of the hypermultiplets. 
\vskip 1.5cm

We observe that our investigation is stricly related to the 
``non--chiral'' analogue of the Picard--Fuchs equations 
\cite{is,picardfucses} and the related integrable structure
\cite{integrable}.  Also, the approach deserves to be generalised to 
the higher--rank group case \cite{generalizations,related} and to
the strong coupling region \cite{BilalFerrari}. 
Finally, we observe that much of the theory seems related to
Duistermaat--Heckman theorem \cite{DuistermaatHeckman}.
 In this context we observe that
in a recent paper
McArthur and Gargett 
a ``Gaussian approach'' to supersymmetric effective actions
has been investigated \cite{McArthurGargett}.

\vskip 3cm
\leftline{\bf\large Acknowledgements}
\vskip 1.5cm
\noindent
D.~B. would like to thank Marc Grisaru for discussions 
on the papers \cite{DEWIT,ZANON}. 
F.~F. was partly supported by 
the European Commission Contract CHRX--CT93--0340 and 
M.~M. by the European Commission TMR programme ERBFMRX--CT96--0045.


\newpage
\begin{thebibliography}{99}

\bibitem{sw}
{N.~Seiberg and E.~Witten, Nucl. Phys. {\bf B426} (1994) 19;
{\it ibid.} {\bf B430} (1994) 485.}

\bibitem{sw2}
{N.~Seiberg and E.~Witten, Nucl. Phys. {\bf B431} (1994) 484.}


\bibitem{mat2}
{G.~Bonelli,  M.~Matone  and M.~Tonin,
Phys. Rev. {\bf D55} (1997) 6466.}

\bibitem{mat}
{M.~Matone, Phys. Lett. {\bf 357B} (1995) 342.}

\bibitem{sonne}   
J. Sonnenschein,  S. Theisen and  S. Yankielowicz,     
Phys. Lett. {\bf B367} (1996) 145; 
T. Eguchi and S.-K. Yang, Mod. Phys. Lett. {\bf A11} (1996) 131.


\bibitem{fp}
{D.~Finnell and P.~Pouliot, Nucl. Phys. {\bf B453} (1995) 225.}

\bibitem{is}
{K.~Ito and N.~Sasakura, Phys. Lett. {\bf B382} (1996) 95.

\bibitem{fp2}
T. Harano and M. Sato, Nucl. Phys. {\bf B484} (1997) 167; 
H. Aoyama, T. Harano, M. Sato and S. Wada, Phys.Lett. 
{\bf B388} (1996) 331; 
N. Dorey, V. V. Khoze and M. P. Mattis, Phys. Lett. {\bf B388} (1996) 324.}

\bibitem{dkm}
{N.~Dorey, V. V. Khoze  and   M. P. Mattis,  Phys. Rev.  {\bf D54}  (1996)
2921.}

\bibitem{dkm2}
{N.~Dorey, V. V. Khoze  and  M. P. Mattis,  Phys. Lett.  {\bf B390}  (1997)
205.}

\bibitem{DKMMATT}
{N.~Dorey, V. V. Khoze  and  M. P. Mattis,  Phys. Rev.  {\bf D54}  (1996)
7832.}

\bibitem{GAB}
{F.~Fucito  and   G.~Travaglini, 
Phys. Rev. {\bf D55} (1997) 1099.}

\bibitem{HW}
{P. S.~Howe   and  P. C.~West, Nucl. Phys. {\bf B486} (1996) 425.} 
   
\bibitem{HENN} M. Henningson, Nucl. Phys. {\bf B458} (1996) 445.

\bibitem{DEWIT} B. de Wit, M. T. Grisaru and M. Ro\v{c}ek, 
Phys. Lett. {\bf B374} (1996) 297. 

\bibitem{YUNG}  A. Yung, Nucl. Phys. {\bf B485} (1997) 38.

\bibitem{MATK}
M.~Matone, Phys. Rev. Lett. {\bf 78} (1997) 1412.

\bibitem{SD} M. Dine and N.
Seiberg, 
{\it Comment on Higher Derivative Operators in Some SUSY Field Theories},
SCIIP 97/12, RU--97--13, {\tt hep-th/9705057}.

\bibitem{DKMNEW}
N. Dorey, V. V. Khoze, M. P. Mattis, M. J. Slater and W. A. Weir, 
{\it Instantons, Higher Derivative Terms and Nonrenormalization Theorems 
in Supersymmetric Gauge Theories}, {\tt hep-th/9706007}.

\bibitem{m2} M. Matone, Phys. Rev. {\bf D53} (1996) 7354.


\bibitem{ketov}S. V. Ketov, {\it On the next--to--leading--order
correction to the effective action in N=2 gauge theories},
DESY 97--103, ITP--UH--18/97, {\tt hep-th 9706079}.

\bibitem{BoMa} G. Bonelli and M. Matone, Phys. Rev. Lett. 
{\bf 76} (1996) 4107.

\bibitem{BoMa2} G. Bonelli and M. Matone, Phys. Rev. Lett. 
{\bf 77} (1996) 4712.

\bibitem{ZANON} A. De Giovanni, M. T. Grisaru, M. Ro\v{c}ek, 
R. von Unge and D. Zanon, {\it The N=2 Super Yang--Mills Low--Energy 
Effective Action at Two Loops}, IFUM568--FT, {\tt hep-th/9706013}.

\bibitem{Seibergnonren}N. Seiberg,
 Phys. Lett.  {\bf B328} (1993) 469.

\bibitem{gomez}C. Gomez and R. Hernandez, {\it Electric--magnetic duality and
effective field theories}, Madrid preprint FTUAM 96/36, {\tt hep-th/9510023}.



\bibitem{Witten1}E. Witten,
 Nucl. Phys. {\bf B202} (1982) 253.
 
\bibitem{Gates} S. J. Gates, Jr. Nucl. Phys. {\bf B238} (1984) 349.

\bibitem{WB}
{J.~Wess and J.~Bagger, {\it Supersymmetry and Supergravity}, Princeton 
University Press, 1983.}

\bibitem{th}
{G.~'t~Hooft, Phys. Rev. {\bf D14} (1976) 3432.}

\bibitem{sei}
{N.~Seiberg, Phys. Lett. {\bf 206B} (1988) 75.}


\bibitem{gt}
{G. Travaglini, Ph. D. Thesis, Universit\`a di Roma ``La Sapienza", 1996,
unpublished.}

\bibitem{algom}L. Alvarez--Gaum\'e and S. F. Hassan, 
{\it Introduction to S--Duality in $N=2$ Supersymmetric Gauge Theories 
(A Pedagogical Review of the Work of Seiberg and Witten)} CERN--TH/96--371, 
{\tt hep-th/9701069}.


\bibitem{divecchia}
{A.~D'Adda  and P.~Di Vecchia, Phys. Lett. 
{\bf 73B} (1978) 162.}

\bibitem{ber}
{C.~Bernard, Phys. Rev. {\bf D19} (1979) 3013.}


\bibitem{nsvz2}
{V.~Novikov, M.~Shifman, A.~Vainshtein and 
V.~Zakharov, Nucl. Phys. {\bf B223} (1983) 445.}

\bibitem{YUNGQCD}
A. Yung, {\it Higher Derivative Terms 
in the Effective Action of N=2 SUSY QCD from 
Instantons}, {\tt hep-th/9705181}, PNPI--2168-TH--25.


\bibitem{ADHM}{M.~Atiyah, V.~Drinfeld, N.~Hitchin and Yu.~Manin, 
Phys. Lett. 
{\bf 65A} (1978) 185.}

\bibitem{aty}{M.~Atiyah, {\it Geometry of Yang--Mills 
Fields}, Lezioni 
Fermiane, Accademia Nazionale dei Lincei e Scuola Normale 
Superiore, Pisa 
(1979).}

\bibitem{osb}
{H.~Osborn, Ann. Phys. {\bf 135} (1981) 373.}

\bibitem{csw}{N.~Christ, N.~Stanton and  E.~Weinberg, 
 Phys. Rev. {\bf D18} (1978) 2013.}

\bibitem{picardfucses}
A. Ceresole, R. D'Auria and S. Ferrara,
Phys. Lett. {\bf B339} (1994) 71; 
J. M. Isidro, A. Mukherjee,
J. P. Nunes and H. J. Schnitzer, Nucl. Phys. {\bf B492} (1997) 647; 
{\tt hep-th/9703176}; {\tt hep-th/9704174}.
 
\bibitem{generalizations}
A. Klemm, W. Lerche, S. Yankielowicz and S. Theisen,
Phys. Lett. {\bf B344} (1995) 169; P. C. Argyres and A. E. Faraggi, 
Phys. Rev. Lett. {\bf 74} (1995) 3931.

\bibitem{related}
I. M. Krichever and D. H. Phong,
{\it On the Integrable Geometry of Soliton Equations and $N=2$
Supersymmetric Gauge Theories}, {\tt hep-th/9604199};
E. D'Hoker, I. M. Krichever and D. H. Phong, Nucl. Phys. {\bf B489} (1997) 
179; Nucl. Phys. {\bf B489} (1997) 211; 
{\it The Renormalization Group Equation in $N=2$ Supersymmetric Gauge 
Theories}, UCLA--96--TEP--40, {\tt hep-th/9610156};
E. D'Hoker, Phys. Lett. {\bf B397} (1997) 94.

\bibitem{integrable}
A. Gorsky, I. Krichever, A. Marshakov, A. Morozov
and A. Mironov, Phys. Lett. {\bf B355} (1995) 466; R. Donagi and E. 
Witten, Nucl. Phys. {\bf B460} (1996) 299;
E. Martinec and N. P. Warner, Nucl. Phys. {\bf B459} (1996) 97; 
E. Martinec, Phys. Lett. {\bf B367} (1996) 91;
T. Nakatsu and K. Takasaki, Mod. Phys. Lett. {\bf A11} (1996) 157;
Int. J. Mod. Phys. {\bf A11} (1996) 5505; {\tt hep-th/9603219};
H. Itoyama and A. Morozov, Nucl. Phys. {\bf B477} (1996) 855; {\bf B491} 
(1997) 529; {\tt hep-th/9601168}; 
 R. Carroll,  {\tt solv-int/9511009}; 
{\tt solv-int/9606005}.

\bibitem{BilalFerrari} 
F. Ferrari and A. Bilal,
Nucl. Phys. {\bf B469} (1996) 387;
A. Bilal and F. Ferrari, Nucl. Phys. {\bf B480} (1996) 589.

\bibitem{DuistermaatHeckman} J. J. Duistermaat and G. J. Heckman,
Invent. Math. {\bf 69} (1982) 259; R. F. Picken, J. Math. Phys. {\bf 31} 
(1990) 616.


\bibitem{McArthurGargett}
I. N. McArthur and  T. D. Gargett, 
{\it A ``Gaussian" Approach to Computing Supersymmetric Effective Actions}, 
LMU--TPW--97--9, {\tt hep-th/9705200}.


\end{thebibliography}
\end{document}